\documentclass{kluwer}    

\newdisplay{guess}{Conjecture}
\usepackage{epsfig}
\usepackage{graphicx}

\def\thefigure{\arabic{figure}}

\begin{document}
\begin{article}
\begin{opening}

{\title{The geometry of density states, positive maps\\ and tomograms}
\author{V.I. Man'ko,$^1$ G. Marmo,$^2$ E.C.G. Sudarshan$^3$ and
F. Zaccaria$^2$}
\institute{$^1$P.N. Lebedev Physical Institute, Leninskii Prospect,
53, Noscow 119991 Russia\\$^2$Dipartimento di Scienze Fisiche,
Universit\`{a} ``Federico II'' di Napoli and Istituto Nazionale di
Fisica Nucleare, Sezione di Napoli, Complesso Universitario di Monte
Sant Angelo, Via Cintia, I-80126 Napoli, Italy\\$^3$Physics
Department, Center for Particle Physics, University of Texas, 78712
Austin, Texas, USA\\ Emails {\small \bf manko@sci.lebedev.ru~~
marmo@na.infn.it\\sudarshan@physics.utexas.edu~~zaccaria@na.infn.it }
}}
\runningauthor{V.I. Man'ko, G. Marmo, E.C.G. Sudarshan and F.
Zaccaria}
\runningtitle{The geometry of density states, positive maps and tomograms}

\begin{abstract}
The positive and not completely positive maps of density matrices,
which are contractive maps, are discussed as elements of a semigroup.
A new kind of positive map (the purification map), which is nonlinear
map, is introduced. The density matrices are considered as vectors,
linear maps among matrices are represented by superoperators given in
the form of higher dimensional matrices. Probability representation of
spin states (spin tomography) is reviewed and $U(N)$-tomogram of spin
states is presented. Properties of the tomograms as probability
distribution functions are studied. Notion of tomographic purity of
spin states is introduced. Entanglement and separability of density
matrices are expressed in terms of properties of the tomographic joint
probability distributions of random spin projections which depend also
on unitary group parameters. A new positivity criterion for hermitian
matrices is formulated. An entanglement criterion is given in terms of
a function depending on unitary group parameters and semigroup of
positive map parameters. The function is constructed as sum of moduli
of $U(N)$-tomographic symbols of the hermitian matrix obtained after
action on the density matrix of composite system by a positive but not
completely positive map of the subsystem density matrix. Some
two-qubit and two-qutritt states are considered as examples of
entangled states. The connection with the star-product quantisation is
discussed. The structure of the set of density matrices and their
relation to unitary group and Lie algebra of the unitary group are
studied. Nonlinear quantum evolution of state vector obtained by means
of applying purification rule of density matrices evolving via
dynamical maps is considered. Some connection of positive maps and
entanglement with random matrices is discussed and used.
\end{abstract}


\end{opening}

\section{Introduction}

The states in quantum mechanics are associated with vectors in Hilbert
space~[1]
(it is better to say with rays) in the case of pure states. For mixed
state, one associates the state with density matrix~[2,~3].
In classical mechanics (statistical mechanics), the states are
associated with joint probability distributions in phase space. There
is an essential difference in the concept of states in classical and
quantum mechanics. This difference is clearly pointed out by the
phenomenon of entanglement. The notion of entanglement~[4]
is related to the quantum superposition rule of the states of
subsystems for a given multipartite system. For pure states, the
notion of entanglement and separability can be given as follows.

If the wave function~[5]
of a state of a bipartite system is represented as the product of two
wave functions depending on coordinates of the subsystems, the state
is simply separable; otherwise, the state is simply entangled. An
intrinsic approach to the entanglement measure was suggested in [6].
The measure was introduced as the distance between the system density
matrix and the tensor product of the associated states. For the
subsystems, the association being realized via partial traces. There
are several other different characteristics and measures of
entanglement considered by several authors [7--13].
For example, there are measures related to entropy (see, [14--24]).
Also linear entropy of entanglement was used in [25--27],
``concurrences'' in [28,~29] and ``covariance entanglement measure''
in [30]. Each of the entanglement measures describes some degree of
correlation between the subsystems' properties.

The notion of entanglement is not an absolute notion for a given
system but depends on the decomposition into subsystems. The same
quantum state can be considered as entangled, if one kind of division
of the system into subsystems is given, or as completely disentangled,
if another decomposition of the system into subsystems is considered.

For instance, the state of two continuous quadratures can be entangled
in Cartesian coordinates and disentangled in polar coordinates.
Coordinates are considered as measurable observables labelling the
subsystems of the given system. The choice of different subsystems
mathematically implies the existence of two different sets of the
subsystems' characteristics (we focus on bipartite case). We may
consider the Hilbert space of states $H(1,2)$ or $H(1^{\prime
},2^{\prime })$. The  Hilbert space for the total system is, of
course, the same but the index $(1,2)$ means that there are two sets
of operators $P_{1}$ and $P_{2}$, which select subsystem states 1  and
2. The index $(1^{\prime },2^{\prime })$ means that there are two
other sets of operators $P_{1}^{\prime }$ and  $P_{2}^{\prime }$,
which select subsystem states $1^{\prime }$ and $2^{\prime }.$  The
operators $P_{1,2}$ and $P_{1^{\prime },2^{\prime }}^{\prime }$ have
specific properties. They are represented as tensor products of
operators acting in the space of states of the subsystem 1 (or 2) and
unit operators acting in the subsystem 2 (or 1). In other words, we
consider the space $H$, which can be treated as the tensor product of
spaces  $H(1)$ and $H(2)$ or $H(1')$ and $H(2')$. In the subsystems
$1$ and $2$, there are basis vectors $\mid n_{1}\rangle$ and $ \mid
m_{2}\rangle$, and in the subsystems $1^{\prime }$ and  $2^{\prime }$
there are basis vectors $\mid n_{1}^{\prime }\rangle$ and  $\mid
m_{2}^{\prime }\rangle.$ The vectors  $\mid n_{1}\rangle
\mid m_{2}\rangle$ and $\mid n_{1}^{\prime }
\rangle\mid m_{2}^{\prime}\rangle$ form the sets
of basis vectors in the composite Hilbert space, respectively. These
two sets are related by means of unitary transformation. An example of
such a composite system is a bipartite spin system.

If one has spin-$j_{1}$ [the space $H(1)$] and spin-$j_{2}$ [the space
$H(2)$] systems, the combined system can be treated as having basis\\
$\mid j_{1}m_{1}\rangle\mid j_{2}m_{2}\rangle.$

Another basis in the composite-system-state space can be considered in
the form  $\mid j_{{}}m_{{}}\rangle$, where $j$ is one of the numbers
$|j_{1}-j_{2}|,$ $|j_{1}-j_{2}|+1,\ldots,$ $j_{1}+j_{2}$ and
$m=m_{1}+m_{2}$. The basis $\mid jm\rangle$ is related to the basis
$\mid j_{1}m_{1}\rangle\mid j_{2}m_{2}\rangle$ by means of the unitary
transform given by Clebsch--Gordon coefficients $C$ $(j_{1}m_{1}
j_{2}m_{2}|jm)$. From the viewpoint of the given definition, the
states $\mid j_{{}}m_{{}}\rangle$ are entangled states in the original
basis. Another example is the separation of the hydrogen atom in terms
of parabolic coordinates used while discussing the Stark effect.

The spin states can be described by means of the tomographic
map~[31--33].
For bipartite spin systems, the states were described  by the
tomographic probabilities in [34,~35].
Some properties of the tomographic spin description were studied in
[36].
In the tomographic approach, the problems of the quantum state
entanglement can be cast into the form of some relations among the
probability distribution functions. On the other hand, to have a clear
picture of entanglement, one needs a mathematical formulation of the
properties of the density matrix of the composite system, a
description of the linear space of the composite system states. Since
a density matrix is hermitian, the space of states may be embedded as
a subset of the Lie algebra of the unitary group, carrying the adjoint
representation of $U(n^{2})$, where $n^2=(2j+1)^2$ is the dimension of
the spin states of two spinning particles. Thus one may try to
characterize the entanglement phenomena by using various structures
present in the space of the adjoint representation of the $U(n^{2})$
group.

The aim of this paper is to give a review of different aspects of
density matrices and positive maps and connect entanglement problems
with the properties of tomographic probability distributions and
discuss the properties of the convex set of positive states for
composite system by taking into account the subsystem structures. We
used~[6]
the Hilbert--Schmidt distance to calculate the measure of entanglement
as the distance between a given state and the tensor product of the
partial traces of the density matrix of the given state. In [37]
another measure of entanglement as a characteristic of subsystem
correlations was introduced. This measure is determined via the
covariance matrix of some observables. A review of different
approaches to the entanglement notion and entanglement measures is
given in [38]
where the approach to describe entanglement and separability of
composite systems is based, e.g., on entropy methods.

Due to a variety of approaches to the entanglement problem, one needs
to understand better what in reality the word ``entanglement''
describes. Is it a synonym of the word ``correlation'' between two
subsystems or does it have to capture some specific correlations
attributed completely and only to the quantum domain?

The paper is organized as follows.

In section~2 we discuss division of composite systems onto subsystems
and relation of the density matrix to adjoint representation of
unitary group in generic terms of vector representation of matrices;
we study also completely positive maps of density matrices. In
section~3 we consider a vector representation of probability
distribution functions and notion of distance between the probability
distributions and density matrices. In section~4 we present definition
of separable quantum state of a composite system and criterion of
separability. In section~5 the entanglement is considered in terms of
operator symbols. Particular tomographic probability representation of
quantum states and tomographic symbols are reviewed in section~6.
Symbols of multipartite states are studied in section~7. In section~8
spin tomography is reviewed. An example of qubit state is done in
section~9. The unitary spin tomogram is introduced in section~10 while
in section~11 dynamical map and corresponding quantum evolution
equations are discussed as well as examples of concrete positive maps.
Conclusions and perspectives are presented in section~12.

\section{Composite system }

In this section, we review the meaning and notion of composite
system in terms of additional structures on the linear space of
state for the composite system.

\subsection{Difference of states and observables}

In quantum mechanics, there are two basic aspects, which are
associated with linear operators acting in a Hilbert space. The first
one is related to the concept of quantum state and the second one, to
the concept of observable. These two concepts of state and observable
are paired via a map with values in probability measures on the real
line. Often states are described by Hermitian nonnegative,
trace-class, matrices. The observables are described by Hermitian
operators. Though both states and observables are identified with
Hermitian objects, there is an essential difference between the
corresponding objects. The observables have an additional product
structure. Thus we may consider the product of two linear operators
corresponding to observables.

For the states, the notion of product is redundant. The product of two
states is not a state. For states, one keeps only the linear structure
of vector space. For finite $n$-dimensional system, the Hermitian
states and the Hermitian observables may be mapped into the Lie
algebra of the unitary group $U(n)$. But the states correspond to
nonnegative Hermitian operators. Observables can be associated with
both types of operators, including nonnegative and nonpositive ones.
The space of states is therefore a convex-linear space which, in
principle, is not equipped with a product structure. Due to this,
transformations in the linear space of states need not preserve any
product structure. In the set of observables, one has to be concerned
with what is happening with the product of operators when some
transformations are performed. State vectors can be transformed into
other state vectors. Density operators also can be transformed. We
will consider linear transformations of the density operators. The
density operator has nonnegative eigenvalues. In any representation,
diagonal elements of density matrix have physical meaning of
probability distribution function.

Density operator can be decomposed as a sum of eigenprojectors with
coefficients equal to its eigenvalues. Each one of the projectors
defines a pure state. There exists a basis in which every
eigenprojector of rank one is represented by a diagonal matrix of rank
one with only one matrix element equal to one and all other matrix
elements equal to zero. Other density matrices with similar properties
belong to the orbit of the unitary group on the starting
eigenprojector. Depending on the number of distinct nonzero values
determines the class of the orbit. Since density matrices of higher
rank belong to an appropriate orbit of a convex sum of the different
diagonal eigenprojectors (in special basis), we may say that generic
density matrices belong to the orbits of the unitary group acting on
the diagonal density matrices which belong to the Cartan subalgebra of
the Lie algebra of the unitary group. Any convex sum of density
matrices can be treated as a mean value of a random density matrix.
The positive coefficients of the convex sum can be interpreted as a
probability distribution function which makes the averaging providing
the final value of the convex sum. The set of density matrices may be
identified with the union of the orbits of the unitary group acting on
diagonal density matrices considered as elements of the Cartan
subalgebra.

\subsection{Matrices as vectors, density operators and superoperators}

When matrices represent states it may be convenient to identify them
with vectors. In this case, a density matrix can be considered as a
vector with additional properties of its components. If the
identifications are done elegantly, we can see the real Hilbert space
of density matrices in terms of vectors with real components. In this
case, linear transforms of the matrix can be interpreted as matrices
called superoperators. It means that density matrices--vectors
undergoing real linear transformations are acted on by the matrices
representing the action of the superoperators of the linear map.  This
construction can be continued. Thus we can get a chain of vector
spaces of higher and higher dimensions. Let us first introduce some
extra constructions of the map of a matrix onto a vector. Given a
rectangular matrix $M$ with elements $M_{id}$, where $i=1,2,\ldots,n$
and $d=1,2,\ldots,m$, one can consider the matrix as a vector
$\mathcal{\vec{M}}$ with $N=nm$ components constructed by the
following rule:
\begin{eqnarray}
\label{eq.1}
&& \mathcal{M}_{1}=M_{11}, \qquad\mathcal{M}_{2}=M_{12},
\qquad
\mathcal{M}_{m}=M_{1m},\nonumber\\
&&\\ &&\mathcal{M}_{m+1}=M_{21},
\ldots \mathcal{M}_{N}=M_{nm}.\nonumber
\end{eqnarray}
Thus we construct the map $M\rightarrow \mathcal{\vec{M}=}
\hat{t}_{\mathcal{\vec{M}}M}M.$

We have introduced the linear operator $\hat{t}_{\mathcal{\vec{M}}M}$
which maps the matrix $M$ onto a vector $\mathcal{\vec{M}}$. Now we
introduce the inverse operator $\hat{p}_{\mathcal{\vec{M}}M}$ which
maps a given column vector in the space with dimension $N=mn$ onto a
rectangular matrix. This means that given a vector
$\mathcal{\vec{M}=M}_{1},\ldots,\mathcal{M}_{N}$, we relabel its
components by introducing two indices $i=1,\ldots,n$ and
$d=1,\ldots,m.$  The relabeling is accomplished according to
(\ref{eq.1}).  Then we collect the relabeled components into a matrix
table. Eventually we get the map
\begin{equation}\label{eq.2}
\hat{p}_{\mathcal{\vec{M}}M}\mathcal{\vec{M}}=M.
\end{equation}
The composition of these two maps
\begin{equation}\label{eq.3}
\hat{t}_{\mathcal{\vec{M}}M}\hat{p}_{\mathcal{\vec{M}}M}
\mathcal{\vec{M}=}1\cdot\mathcal{\vec{M}}
\end{equation}
acts as the unit operator in the linear space of vectors.

Given a $n$$\times$$n$ matrix the map considered can also be applied.
The matrix can be treated as an $n^{2}$-dimensional vector and, vice
versa, the vector of dimension $n^{2}$ may be mapped by this procedure
onto the $n$$\times$$n$ matrix.

Let us consider a linear operator acting on the vector $\vec{\cal M}$
and related to a linear transform of the matrix $M$. First, we study
the correspondence of the linear transform of the form
\begin{equation}\label{L1}
M\rightarrow gM=M_g^l
\end{equation}
to the transform of the vector
\begin{equation}\label{L2}
\vec{\cal M}\rightarrow \vec{\cal M}_g^l={\cal L}_g^l\vec {\cal M}.
\end{equation}
One can show that the $n^2$$\times$$n^2$ matrix ${\cal L}_g^l$ is
determined by the tensor product of the $n$$\times$$n$ matrix $g$ and
$n$$\times$$n$ unit matrix, i.e.,
\begin{equation}\label{L3}
{\cal L}_g^l=g\otimes 1.
\end{equation}
Analogously, the linear transform of the matrix $ M$ of the form
\begin{equation}\label{L4}
M\rightarrow Mg=M_g^r
\end{equation}
induces the linear transform of the vector $\vec{\cal M}$ of the form
\begin{equation}\label{L5}
\vec{\cal M}\rightarrow \vec{\cal M}_g^r=\hat t_{\vec {\cal M}
{\cal M}}M_g^r={\cal L}_g^r\vec
{\cal M},
\end{equation}
where the $n^2$$\times$$n^2$ matrix ${\cal L}_g^r$ reads
\begin{equation}\label{L6}
{\cal L}_g^r=1\otimes g^{\mbox{tr}}.
\end{equation}
Similarity transformation of the matrix $M$ of the form
\begin{equation}\label{L7}
M\rightarrow gMg^{-1}
\end{equation}
induces the corresponding linear transform of the vector
$\vec{\cal M}$ of the form
\begin{equation}\label{L8}
\vec{\cal M}\rightarrow \vec{\cal M}_s={\cal L}_g^s\vec {\cal M},
\end{equation}
where the $n^2$$\times$$n^2$ matrix ${\cal L}_g^s$ reads
\begin{equation}\label{L9}
{\cal L}_g^s=g\otimes (g^{-1})^{\mbox{tr}}.
\end{equation}
Starting with vectors, one may ask how to identify  on them a product
structure which would make  $\hat p_{\vec{\cal M}{\cal N}}$ into an
algebra homomorphism. An associative algebraic structure on the vector
space may be defined by imitating the procedure one uses to define
star-products on the space of functions on phase space. One can define
the associative product of two $N$-vectors $\vec{\cal M}_1$ and
$\vec{\cal M}_2$ using the rule
\begin{equation}\label{L10}
\vec{\cal M}=\vec{\cal M}_1\star\vec{\cal M}_2,
\end{equation}
where
\begin{equation}\label{L11}
\vec{\cal M}_k=\sum_{l,s=1}^NK_{ls}^k(\vec{\cal M}_1)_l(\vec{\cal M}_2)_s.
\end{equation}
If one applies a linear transform to the vectors $\vec{\cal M}_1$,
$\vec{\cal M}_2$, $\vec{\cal M}$ of the form
$$ \vec{\cal M}_1\rightarrow\vec{\cal M}^\prime_1= {\cal
L}\vec{\cal M}_1,\qquad \vec{\cal M}_2\rightarrow\vec{\cal M}^\prime_2
= {\cal L}\vec{\cal M}_2,\qquad \vec{\cal M}\rightarrow\vec{\cal M}^\prime
= {\cal L}\vec{\cal M}, $$
and requires the invariance of the star-product kernel, one finds
$$
\vec{\cal M}_1^\prime \star\vec{\cal M}^\prime_2= \vec{\cal
M}^\prime,\qquad \mbox{if}\qquad {\cal L}=G\otimes
G^{-1\mbox{tr}},\quad G\in GL(n).$$ The kernel $K_{ls}^k$ (structure
constants) which determines the associative star-product satisfies a
quadratic equation. Thus if one wants to make the correspondence of
the vector star-product to the standard matrix product (row by
column), the matrix $M$ must be constructed appropriately. For
example, if the vector star-product is commutative, the matrix $M$
corresponding to the $N$-vector $\vec{\cal M}$ can be chosen as a
diagonal $N$$\times$$N$ matrix. This consideration shows that the map
of matrices on the vectors provides the star-product of the vectors
(defining the structure constants or the kernel of the star-product)
and, conversely, if one starts with vectors and uses matrices with the
standard multiplication rule, it will be the map to be determined by
the structure constants (or by the kernel of the vector star-product).

The constructed space of matrices associated with vectors enables one
to enlarge the dimensionality of the group acting in the linear space
of matrices in comparison with the standard one, i.e., we may relax
the requirement of invariance of the product structure. In general,
given a $n$$\times$$n$ matrix $M$ the left action, the right action,
and the similarity transformation of the matrix are related to the
complex group $GL(n)$. On the other hand, the linear transformations
in the linear space of $n^2$-vectors $\vec{\cal M}$ obtained by using
the introduced map are determined by the matrices belonging to the
group $GL(n^2)$. There are transformations on the vectors which cannot
be \underline{simply} represented on matrices. If $M\to\Phi(M)$ is a
linear homogeneous function of the matrix $M$, we may represent it by
$$\Phi_{ab}=B_{aa',\,bb'}M_{a'b'}.$$ Under rather clear conditions,
$B_{aa',bb'}$ can be expressed in terms of its nonnormalized left and
right eigenvectors:
$$B_{aa',bb'}=\sum_\nu x_{aa'}(\nu)y^\dagger_{bb'}(\nu), $$
being an index for eigenvalues, which corresponds to $$
\Phi(M)=xMy^\dagger=\sum_{\nu=1}^{n^2}x(\nu) My^\dagger(\nu).$$

There are possible linear transforms on the matrices and corresponding
linear transforms on the induced vector space which do not give rise
to a group structure but possess only the structure of algebra. One
can describe the map of $n$$\times$$n$ matrices $M$ (source space)
onto vectors $\vec{\cal M}$ (target space) using specific basis in the
space of the matrices. The basis is given by the matrices
$E_{jk}~(j,k=1,2,\ldots,n)$ with all matrix elements equal to zero
except the element in the $j$th row and $k$th column which is equal to
unity. One has the obvious property
\begin{equation}\label{L12}
M_{jk}=\mbox{Tr}\left(ME_{jk}\right).
\end{equation}
In our procedure, the basis matrix $E_{jk}$ is mapped onto the basis
column-vector $\vec{\cal E}_{jk}$, which has all components equal to
zero except the unity component related to the position in the matrix
determined by the numbers $j$ and $k$. Then one has
\begin{equation}\label{L13}
\vec{\cal M}=\sum_{j,k=1}^n\mbox{Tr}\left(ME_{jk}\right)\vec{\cal E}_{jk}.
\end{equation}
For example, for similarity transformation of the finite matrix $M$,
one has
\begin{equation}\label{L14}
\vec{\cal M}_g^s= \sum_{j,k=1}^N\mbox{Tr}\left(gMg^{-1}E_{jk}\right)
\vec{\cal E}_{jk}.
\end{equation}

Now we will define the notion of `composite' vector which corresponds
to dividing a quantum system into subsystems.

We will use the following terminology.

In general, the given linear space of dimensionality $N=mn$ has a
structure of a bipartite system, if the space is equipped with the
operator $\hat{p}_{\mathcal{\vec{M}}M}$ and the matrix $M$ (obtained
by means of the map) has matrix elements in factorizable form
\begin{equation}\label{eq.4}
M_{id}\rightarrow x_{i}y_{d}.
\end{equation}
This $M=x\otimes y$ corresponds to the special case of nonentangled
states. Otherwise, one needs
$$M=\sum_{\nu}x(\nu)\otimes y(\nu).$$ In fact, to consider in
detail the entanglement phenomenon, in the bipartite system of
spin-1/2, one has to introduce a hierarchy of three linear spaces. The
first space of pure spin states is the two-dimensional linear space of
complex vectors
\begin{equation}\label{new1a}
\mid \vec x\rangle=\left(
\begin{array}{c}
x_1 \\x_2
\end{array}
\right).  \end{equation}
In this space, the scalar product is defined as follows:
\begin{equation}\label{new2a}
\langle \vec x\mid\vec y\rangle=x_1^*y_1+x_2^*y_2.
\end{equation}
So it is a two-dimensional Hilbert space. We do not equip this space
with a vector star-product structure. In the primary linear space, one
introduces linear operators $\hat M$ which are described by 2$\times$2
matrices $M$. Due to the map discussed in the previous section, the
matrices are represented by 4-vectors $\vec{\cal M}$ belonging to the
second complex 4-dimensional space. The star-product of the vectors
$\vec{\cal M}$ determined by the kernel ${\cal K}_{ls}^k$ is defined
in such a manner in order to correspond to the standard rule of
multiplication of the matrices.

In addition to the star-product structure, we introduce the scalar
product of the vectors $\vec{\cal M}_1$ and $\vec{\cal M}_2$, in view
of the definition
\begin{equation}\label{new3a}
\langle \vec {\cal M}_1\mid\vec {\cal M}_2\rangle =\mbox{Tr}\,
(M_1^\dagger M_2),
\end{equation}
which is the trace formula for the scalar product of matrices.

This means introducing the real metric $g^{\alpha\beta}$ in the
standard notation for scalar product
\begin{equation}\label{new4a}
\langle \vec {\cal M}_1\mid\vec {\cal M}_2\rangle =\sum_{\alpha,
\beta=1}^4({\cal M}_1)^*_\alpha g^{\alpha\beta}({\cal M}_2)_\beta,
\end{equation}
where the matrix $g^{\alpha\beta}$ is of the form
\begin{equation}\label{new5a}
g^{\alpha\beta}=\left(
\begin{array}{clcr}
1&0&0&0 \\
0&0&1&0 \\
0&1&0&0 \\
0&0&0&1
\end{array}
\right),\qquad g^{\alpha j}g^{j\beta}=\delta^{\alpha\beta}.
\end{equation} The scalar product is invariant under the action of
the group of nonsingular 4$\times$4 matrices $\ell$, which satisfy the
condition
\begin{equation}\label{L22}
 g=\ell^\dagger g\ell. \end{equation} The product of matrices $\ell$
satisfies the same condition since $g^2=1.$

Thus, the space of operators $\hat M$ in the primary two-dimensional
space of spin states is mapped onto the linear space which is equipped
with a scalar product (metric Hilbert space structure) and an
associative star-product (kernel satisfying the quadratic
associativity equation). In the linear space of the 4-vectors $\vec
{\cal M}$, we introduce linear operators (superoperators), which can
be associated with the algebra of 4$\times$4 complex matrices.

Let us now focus on density matrices. This means that our matrix $M$
is considered as a density matrix $\rho$ which describes a quantum
state. We consider here the action of the unitary transformation
$U(n)$ of the density matrices and corresponding transformations on
the vector space. If one has the structure of a bipartite system, we
also consider the action of local gauge transformation both in the
``source space'' of density matrices and in the ``target space'' of
the corresponding vectors.

The $n$$\times$$n$ density matrix $\rho $ has matrix elements
\begin{equation}\label{eq.5}
\rho _{ik}=\rho _{ki}^{*}, \qquad\mbox{Tr}\,\rho _{{}}=1, \qquad
\langle\psi |\rho |\psi\rangle \geq 0.
\end{equation}
Since the density matrix is hermitian, it can always be identified as
an element of the convex subset of the linear space associated with
the Lie algebra of $U(n)$ group, on which the group $U(n)$ acts with
the adjoint representation
\begin{equation}\label{eq.6}
\rho \rightarrow \rho _{U}=U\rho U_{{}}^{\dagger }.\end{equation}
The system is said to be bipartite if the space of representation is
equipped with an additional structure. This means that for
$$n^{2}=n_{1}^{{}}\cdot n_{2}^{{}},$$ where, for simplicity,
$n_{1}=n_{2}=n$, one can make first the map of $n$$\times$$n$ matrix
$\rho _{{}}$ onto $n^{2}$-dimensional vector $\vec{\rho}$ according to
the previous procedure, i.e., one equips the space by an operator
$\hat{t}_{\vec{\rho}\rho }$. Given this vector one makes a relabeling
of the vector $\vec{\rho}$ components according to the rule
\begin{equation}\label{eq.7}
\vec{\rho}\rightarrow \rho _{id,ke}, \quad i,k=1,2,\ldots,n_{1},
\quad d,e=1,2,\ldots,n_{2},
\end{equation} i.e., obtaining again the quadratic matrix
\begin{equation}\label{eq.8}
\rho _{q}=\hat{p}_{\rho _{q}\vec{\rho}}\vec{\rho}. \end{equation}
The unitary transform (\ref{eq.6}) of the density matrix induces a
linear transform of the vector $\vec\rho$ of the form
\begin{equation}\label{new1}
\vec\rho\rightarrow \vec\rho _U=(U\otimes U^*)\vec\rho.
\end{equation}
There exist linear transforms (called positive maps) of the density
matrix, which preserve its trace, hermicity, and positivity. In some
cases, they have the following form introduced in [39]
\begin{equation}\label{new2}
\rho_0\rightarrow \rho _U=L_U\rho_0=\sum_kp_kU_k\rho_0U_k^\dagger,
\qquad\sum_kp_k=1,
\end{equation}
where $U_k$ are unitary matrices and $p_k$ are positive numbers.

If the initial density matrix is diagonal, i.e., it belongs to the
Cartan subalgebra of the Lie algebra of the unitary group, the
diagonal elements of the obtained matrix give a ``smoother''
probability distribution than the initial one. A generic
transformation preserving previously stated properties may be given in
the form (see [39,~40])
\begin{equation}\label{new3}
\rho_0\rightarrow \rho_V=L_V\rho_0=\sum_k V_k\rho_0V_k^\dagger,
\qquad\sum_kV_k^\dagger V_k=1.
\end{equation}
For example, if $V_k~(k=1,2,\ldots ,N)$ are taken as square roots of
orthogonal projectors onto complete set of $N$ state, the map provides
the map of the density matrix $\rho_0$ onto diagonal density matrix
$\rho_{0d}$ which has the same diagonal elements as $\rho_0$ has. In
this case, the matrices $V_k$ have the only nonzero matrix element
which is equal to one. Such a map may be called ``decoherence map''
because it removes all nondiagonal elements of the density matrix
$\rho_0$ killing any phase relations. In quantum information
terminology, one uses also the name ``phase damping channel.'' More
general map may be given if one takes $V_k$ as $N$ generic diagonal
density matrices, in which eigenvalues are obtained by $N$ circular
permutations from the initial one. Due to this map, one has a new
matrix with the same diagonal matrix elements but with changed
nondiagonal elements. The purity of this matrix is smaller then the
purity of the initial one. This means that the map is contractive. All
matrices with the same diagonal elements up to permutations belong to
a given orbit of the unitary group.

For a large number of terms with randomly chosen matrices $V_k$ in the
sum in (\ref{new3}), the above map gives the most stochastic density
matrix
$$\rho_0\rightarrow
\rho_s=L_1\rho_0 =(n)^{-1}1.$$ Its four-dimensional matrix $L_1$ for the
qubit case has four matrix elements different from zero. These matrix
elements are equal to one. They have the labels $L_{11}$, $L_{14}$,
$L_{41}$, $L_{44}$. The map with two nonzero matrix elements
$L_{41}=L_{44}=0$ provides pure-state density matrix from any
$\rho_0$. The transform (\ref{new2}) is the partial case of the
transform (\ref{new3}). We discuss the transforms separately since
they are used in the literature in the presented form.

One can see that the constructed map of density matrices onto vectors
provides the corresponding transforms of the vectors, i.e.,
\begin{equation}\label{new4}
\vec\rho_0\rightarrow \vec\rho_U=\sum_kp_k(U_k\otimes U_k^*)\vec\rho_0
\end{equation}
and
\begin{equation}\label{new5}
\vec\rho_0\rightarrow \vec\rho_V=\sum_k(V_k\otimes V_k^*)\vec\rho_0.
\end{equation}
It is obvious that the set of linear transforms of vectors, which
preserve their properties of being image of density matrices, is
essentially larger than the standard unitary transform of the density
matrices.

Formulae (\ref{new4}) and (\ref{new5}) mean that the positive map
superoperators acting on the density matrix in the vector
representation are described by $n^2$$\times$$n^2$ matrices
\begin{equation}\label{new6}
{\cal L}_U=\sum_kp_k(U_k\otimes U_k^*)
\end{equation}
and
\begin{equation}\label{new7}
{\cal L}_V=\sum_kV_k\otimes V_k^*,
\end{equation}
respectively.

The positive map is called ``noncompletely positive'' if
$${\cal L}=\sum_kV_k\otimes V_k^*-\sum_sv_s\otimes v_s^*,
\qquad \sum_kV_k^\dagger V_k-\sum_sv_s^\dagger v_s=1.$$ This map
is related to a possible ``nonphysical'' evolution of a subsystem.

Formula (\ref{new6}) can be considered in the context of random matrix
representation. In fact, the matrix ${\cal L}_U$ can be interpreted as
the weighted mean value of the random matrix $U_k\otimes U_k^*$. The
dependence of matrix elements and positive numbers $p_k$ on index $k$
means that we have a probability distribution function $p_k$ and
averaging of the random matrix $U_k\otimes U_k^*$ by means of the
distribution function. So the matrix ${\cal L}_U$ reads
\begin{equation}\label{R1}
{\cal L}_U=\langle U\otimes U^*\rangle.
\end{equation}
Let us consider an example of a 2$\times$2 unitary matrix. We can
consider a matrix of the $SU(2)$ group of the form
\begin{equation}\label{R2}
u=\left(
\begin{array}{cl}
\alpha&\beta\\
-\beta^*&\alpha^*
\end{array}
\right),\qquad |\alpha|^2+|\beta|^2=1.  \end{equation}
The 4$\times$4 matrix ${\cal L}_U$ takes the form
\begin{equation}\label{R3}
{\cal L}_U=\left(
\begin{array}{clcr}
\ell&m&m^*&1-\ell\\
-n&s&-q&n\\
-n^*&-q^*&s^*&n^*\\
1-\ell&-m&-m^*&\ell
\end{array}
\right).  \end{equation}
The matrix elements of the matrix ${\cal L}_U$ are the means
\begin{eqnarray}
m&=&\langle \alpha\beta^*\rangle,\nonumber\\ \ell&=&\langle
\alpha\alpha^*\rangle,\nonumber\\
n&=&\langle \alpha\beta\rangle,\label{R4}\\
s&=&\langle \alpha^2\rangle,\nonumber\\
q&=&\langle \beta^2\rangle.\nonumber
\end{eqnarray}
The moduli of these matrix elements are smaller than unity.

The determinant of the matrix ${\cal L}_U$ reads
\begin{equation}\label{R5}
\mbox{det}\,{\cal L}_U=(1-2\ell)\Big(|q|^2-|s|^2\Big)+4\,\mbox{Re}
\,\Big[q^*m^*n+mns^*\Big].
\end{equation}
If one represents the matrix ${\cal L}_U$ in block form
\begin{equation}\label{R6}
{\cal L}_U =\left(
\begin{array}{cl}
A&B\\
C&D\end{array} \right),  \end{equation} then
\begin{equation}\label{R7}
A =\left(
\begin{array}{cl}
\ell&m\\
-n&s\end{array} \right),  \qquad B =\left(
\begin{array}{cl}
m^*&1-\ell\\ -q&n\end{array} \right),  \end{equation} and
\begin{equation}\label{R8}
D=\sigma_2A^*\sigma_2,\qquad C=-\sigma_2B^*\sigma_2,
\end{equation} where $\sigma_2$ is the Pauli matrix.

One can check that the product of two different matrices ${\cal L}_U$
can be cast in the same form. This means that the matrices ${\cal
L}_U$ form a 9-parameter compact semigroup. It means that the product
of two matrices from the set (semigroup) belongs to the same set. It
means that composition is inner like the one for groups. There is a
unity element in the semigroup, however, there exist elements which
have no inverse. In our case, these elements are described, e.g., by
the matrices with zero determinant. Also the elements, which are
matrices with nonzero determinants, have no inverse elements in this
set, since the map corresponding to the inverse of these matrices is
not positive one. For example, in the case $\ell=1/2$ and $m=0$, one
has the matrices
\begin{equation}\label{R10}
A =\left(
\begin{array}{cl}
1/2&0\\
-n&s\end{array} \right),  \qquad B =\left(
\begin{array}{cl}
0&1/2\\
-q&n\end{array} \right).  \end{equation}
The determinant of the matrix ${\cal L}_U$ in this case is equal to
zero. All the matrices ${\cal L}_U$ have the eigenvector
\begin{equation}\label{R11}
\vec\rho_0=\left(
\begin{array}{c}
1/2\\0\\0\\1/2\end{array} \right),  \end{equation} i.e.,
\begin{equation}\label{R12}
{\cal L}_U\vec\rho_0=\vec\rho_0.\end{equation} This eigenvector
corresponds to the density matrix
\begin{equation}\label{R13}
\rho_1=\left(
\begin{array}{cl}
1/2&0\\ 0&1/2\end{array} \right),  \end{equation} which is obviously
invariant of the positive map.

For random matrix, one has correlations of the random matrix elements,
e.g., $\langle \alpha\alpha^*\rangle\neq\langle
\alpha\rangle\langle\alpha^*\rangle.$

The matrix ${\cal L}_p$
\begin{equation}\label{R14}
{\cal L}_p=\left(
\begin{array}{clcr}
1&0&0&0\\
0&0&1&0\\
0&1&0&0\\
0&0&0&1
\end{array}
\right)  \end{equation} maps the vector
\begin{equation}\label{R15}
\vec\rho_{\rm in}=\left(
\begin{array}{c}
\rho_{11}\\\rho_{12}\\\rho_{21}\\\rho_{22}\end{array} \right)
\end{equation} onto the vector
\begin{equation}\label{R16}
\vec\rho_{\rm t}=\left(
\begin{array}{c}
\rho_{11}\\\rho_{21}\\\rho_{12}\\\rho_{22}\end{array} \right).
\end{equation} This means that the positive map (\ref{R14})
connects the positive density matrix with its transpose (or complex
conjugate). This map can be presented as the connection of the matrix
$\rho$ with its transpose of the form
$$\rho\to\rho^{\rm tr}=\rho^*=\frac{1}{2}\,\Big(\rho+\sigma_1\rho\sigma_1
-\sigma_2\rho\sigma_2+\sigma_3\rho\sigma_3\Big).$$
There is no unitary transform connecting these matrices.

There is noncompletely positive map in the $N$-dimensional case, which
is given by the generalized formula (for some $\varepsilon$)
$$\rho\rightarrow\rho_s=-\varepsilon
\rho+\frac{1+\varepsilon}{N}\,1_N.$$
In quantum information terminology, it is called ``depolarizing
channel.''

For the qubit case, matrix form of this map reads
\begin{equation}\label{R14*}
L=\left(
\begin{array}{clcr}
\frac{1-\varepsilon}{2}&0&0&\frac{1+\varepsilon}{2}\\
 0&-\varepsilon&0&0\\
0&0&-\varepsilon&0\\
\frac{1+\varepsilon}{2}&0&0&\frac{1-\varepsilon}{2}
\end{array}
\right).  \end{equation}

Thus we constructed the matrix representation of the positive map of
density operators of the spin-1/2 system. This particular set of
matrices realize the representation of the semigroup of real numbers
$-1\leq\varepsilon\leq 1$. If one considers the product
$\varepsilon_1\varepsilon_2=\varepsilon_3$, the result $\varepsilon_3$
belongs to the semigroup. Only two elements 1 and $-1$ of the
semigroup have the inverse. These two elements form the finite
subgroup of the semigroup. The semigroup itself without element
$\varepsilon =0$ can be embedded into the group of real numbers with
natural multiplication rule. Each matrix $L$ has an inverse element in
this group but all the parameters of the inverse elements $\eta$ live
out of the segment $-1,1$. The group of the real numbers is
commutative. The matrices of the nonunitary representation of this
group commute too. It means that we have nonunitary reducible
representation of the semigroup which is also commutative. To
construct this representation, one needs to use the map of matrices on
the vectors discussed in the previous section. Formulae (\ref{new3})
and (\ref{new7}) can be interpreted also in the context of the random
matrix representation, but we use the uniform distribution for
averaging in this case. So one has equality (\ref{new7}) in the form
\begin{equation}\label{R20}
{\cal L}_V=\langle V\otimes V^*\rangle
\end{equation}
and the equality
\begin{equation}\label{R21}
\langle V^\dagger V\rangle=1,
\end{equation}
which provides constraints for the random matrices $V$ used.

Using the random matrix formalism, the positive (but not
completely positive) maps can be presented in the form $${\cal
L}=\langle V\otimes V^*\rangle-\langle v\otimes v^*\rangle,\qquad
\langle V^\dagger V\rangle -\langle v^\dagger v\rangle=1.$$
One can characterize the action of positive map on a density matrix
$\rho$ by the parameter
$$\kappa=\frac{\mbox{Tr}\left({\cal L}\rho\right)^2}{\mbox{Tr}\,\rho^2}
=\frac{\mu_{{\cal L}\rho}}{\mu_\rho}\leq 1.
$$
As a remark we note that in [39]
the positive maps~(\ref{new2}) and (\ref{new3}) were used to describe
the non-Hamiltonian evolution of quantum states for open systems.

We have to point out that, in general, such evolution is not described
by first-order-in-time differential equation. As in the previous case,
if there are additional structures for the matrix in the form
\begin{equation}\label{eq.9}
\rho _{id,ke}\rightarrow x_{i}y_{d}z_{k}t_{e},
\end{equation} which means associating with the initial linear space
two extra linear spaces where $x_{i},z_{k}$ are considered as vector
components in the $n_{1}$-dimensional linear space and $y_{d}$,
$t_{e}$ are vector components in $n_{2}$-dimensional vector space, we
see that one has bipartite structure of the initial space of state
[bipartite structure of the space of adjoint representations of the
group $U(n)$].

Usually the adjoint representation of any group is defined per se
without any reference to possible substructures. Here we introduce the
space with extra structure. In addition to being the space of the
adjoint representation of the group $U(n)$, it has the structure of a
bipartite system. The generalization to multipartite ($N$-partite)
structure is straightforward. One needs only the representation of
positive integer $n^{2}$ in the form
\begin{equation}\label{eq.9a}
n^{2}=\prod_{k=1}^{N}n_{k}^{2}. \end{equation}

If one considers the more general map given by superoperator
(\ref{new7}) rewritten in the form
$${\cal L}_V=\langle V\otimes V^*\rangle,\qquad
\langle V^\dagger V\rangle=1,$$ the number of parameters
determining the matrix ${\cal L}_V$ can be easily evaluated. For
example, for $n=2$,
$$
V=\left(
\begin{array}{clcr}
a&b\\
c&d\end{array} \right),\qquad V^*=\left(
\begin{array}{clcr}
a^*&b^*\\ c^*&d^*\end{array} \right),$$ where the matrix elements are
complex numbers, the normalization condition provides four constraints
for the real and imaginary parts of the matrix elements of the
following matrix: $$ {\cal L}_V=\left(
\begin{array}{clcr}
\langle |a|^2\rangle &\langle ab^*\rangle &
\langle ba^*\rangle &\langle bb^*\rangle\\
\langle ac^*\rangle &\langle ad^*\rangle &
\langle bc^*\rangle &\langle bd^*\rangle\\
\langle ca^*\rangle &\langle cb^*\rangle &
\langle da^*\rangle &\langle db^*\rangle\\
\langle cc^*\rangle &\langle cd^*\rangle &
\langle dc^*\rangle &\langle dd^*\rangle
\end{array}
\right),$$ namely, $$\langle |a|^2\rangle +\langle
|c|^2\rangle=1,\qquad \langle |b|^2\rangle +\langle |d|^2\rangle
=1,\qquad \langle a^*b\rangle +\langle c^*d\rangle=0.$$ Due to the
structure of the matrix ${\cal L}_V$, there are six complex parameters
$$\langle ab^*\rangle, \quad\langle ac^*\rangle,\quad\langle
ad^*\rangle,\quad\langle bc^*\rangle,\quad\langle bd^*\rangle,
\quad\langle cd^*\rangle$$ or 12 real parameters.

The geometrical picture of the positive map can be clarified if one
considers the transform of the positive density matrix into another
density matrix as the transform of an ellipsoid into another
ellipsoid. A generic positive transform means a generic transform of
the ellipsoid, which changes its orientation, values of semiaxis, and
position in the space. But the transform does not change the
ellipsoidal surface into a hyperboloidal or paraboloidal surface. For
pure states, the positive density matrix defines the quadratic form
which is maximally degenerated. In this sense, the ``ellipsoid''
includes all its degenerate forms corresponding to the density matrix
of rank less than $n$ (in $n$-dimensional case). The number of
parameters defining the map $\langle V\otimes V^*\rangle$ in the
$n$-dimensional case is equal to $n^2(n^2-1)$.

The linear space of Hermitian matrices is also equipped with the
commutator structure defining the Lie algebra of the group $U(n)$. The
kernel that defines this structure (Lie product structure) is
determined by the kernel that determines the star-product.

\section{Distributions as vectors}

In quantum mechanics, one needs  the concept of distance between the
quantum states. In this section, we consider the notion of distance
between the quantum states in terms of vectors. First, let us discuss
the notion of distance between conventional probability distributions.
This notion is well known in the classical probability theory.

Given the probability distribution $P(k)$, $k=1,2,\ldots N$, one can
introduce the vector $\vec{P}$ in the form of a column with components
$P_{1}=P(1),$ $P_{2}=P(2),\ldots,$ $P_{N}=P(N).$  The vector satisfies
the condition
\begin{equation}\label{eq.10}
\sum_{k=1}^{N}P_{k}=1. \end{equation} This set of vectors does not form
a linear space but only a convex subset. Nevertheless, in this set one
can introduce a distance between two distributions by using the one
suggested by the vector space structure of the ambient space:
\begin{equation}\label{eq.11}
D^{2}=\left( \vec{P}_{1}-\vec{P}_{2}\right)^{2}=\sum_{k}P_{1k}
P_{1k}+\sum_{k}P_{2k}P_{2k}-2\sum_{k}P_{1k}P_{2k}.\end{equation}

Of course, one may use other identifications of distributions with
vectors.

Since all $ P(k)\geq 0$, one can introduce
$\mathcal{P}_{k}=\sqrt{P(k)}$ as components of the vector
$\mathcal{\vec{P}}$. The $\vec{\cal P}$ can be thought of as a column
with nonnegative components. Then the distance between the two
distributions takes the form
\begin{equation}\label{eq.12}
\mathcal{D}^{2}=\left(\mathcal{\vec{P}}_{1}-\mathcal{\vec{P}}_{2}\right)
^{2}=2-2\sum_{k}\sqrt{P_{1}(k)P_{2}(k)}.\end{equation}

The two different definitions (\ref{eq.10}) and (\ref{eq.11}) can be
used as distances between the distributions.

Let us discuss now the notion of distance between the quantum states
determined by density matrices. In the density-matrix space (in the
set of linear space of the adjoint $U(n)$ representation), one can
introduce distances analogously. The first case is
\begin{equation}\label{eq.13}
\mbox{Tr}\left( \rho _{1}-\rho _{2}\right) ^{2}=D^{2}\end{equation}
and the second case is
\begin{equation}\label{eq.14}
\mbox{Tr}\left( \sqrt{\rho _{1}}-\sqrt{\rho _{2}}\right)^{2}
=\mathcal{D}^{2}.\end{equation}
In fact, the distances introduced can be written naturally as norms of
vectors associated to density matrices
\begin{equation}\label{new21}
D^2=|\vec \rho_1-\vec \rho_2|^2
\end{equation}
and
\begin{equation}\label{new22}
{\cal D}^2=\Big(\vec{(\sqrt \rho_1)}-\vec{(\sqrt \rho_2)}\Big)^2,
\end{equation}
respectively.

In the above expressions, we use scalar product of vectors
${\vec\rho}_1$ and ${\vec\rho}_2$ as well as scalar products of
vectors $\vec{({\sqrt \rho_1})}$ and $\vec{({\sqrt \rho_2})}$,
respectively.

Both definitions immediately follow by identification of either
matrices $\rho _{1}$ and $\rho _{2}$ with vectors according to the map
of the previous sections or matrices $\sqrt{\rho _{1}}$ and $
\sqrt{\rho _{2}}$  with vectors. Since the density matrices $\rho
_{1}$ and $\rho _{2}$ have nonnegative eigenvalues, the matrices
$\sqrt{\rho _{1}}$ and $\sqrt{\rho _{2}}$ are defined without
ambiguity. This means that the vectors $\vec{({\sqrt \rho_1})}$ and
$\vec{({\sqrt \rho_2})}$ are also defined without ambiguity. It is
obvious that using this construction and introducing linear map of
positive vectors $\vec{\sqrt\rho}$, one induces nonlinear map of
density matrices. Other analogous functions, in addition to square
root function, can be used to create other nonlinear positive maps.

\section{Separable systems and separability criterion}

According to the definition, the system density matrix is called
separable (for composite system) but not simply separable, if there is
decomposition of the form
\begin{equation}\label{eq.22}
\rho _{AB}=\sum_{k}p_{k}\Big(\rho _{A}^{(k)}\otimes \rho
_{B}^{(k)}\Big), \qquad \sum_{k}p_{k}=1,\qquad 1\geq p_{k}\geq
0.\end{equation} This is Hilbert's problem of biquadrates. Is a
positive biquadratic the positive sum of products of positive
quadratics? In this formula, one may use also sum over two different
indices. Using another labelling in such sum over two different
indices, this sum can be always represented as the sum over only one
index. The formula does not demand orthogonality of the density
operators $\rho_{A}^{(k)}$ and $\rho _{B}^{(k)}$ for different $k$.
Since every density matrix is a convex sum of pure density matrices,
one could demand that $\rho_A^{(k)}$ and $\rho_B^{(k)}$ be pure. This
formula can be interpreted in the context of random matrix
representation reading
\begin{equation}\label{eq.22'}
\rho_{AB}=\langle\rho_A\otimes \rho_B\rangle,
\end{equation}
where $\rho_A$ and $\rho_B$ are considered as random density matrices
of the subsystems $A$ and $B$, respectively. One can use the clarified
structure of the density matrix set as the union of orbits obtained by
action of the unitary group on projectors of rank one with matrix form
containing only one nonzero  matrix element. Then the separable
density matrix of bipartite composite system  can be always written as
the sum of $n_1n_2$ tensor products (or corresponding mean tensor
product), i.e., in (\ref{eq.22'}) the factors are state projectors.
Each of tensor products contains random unitary matrices of local
transforms of the fixed local projector for one subsystem and for the
second subsystem. It means that an arbitrary projector of rank one of
a subsystem can be always presented in the product form
$\rho_A^{(k)}=u_A^{(k)}\rho_Au_A^{(k)\dagger}$, where $u_A^{(k)}$ is a
unitary local transform  and $\rho_A$ is a fixed projector.

There are several criteria for the system to be separable. We suggest
in the next sections a new approach to the problem of separability and
entanglement based on the tomographic probability description of
quantum states. The states which cannot be represented in the form
(\ref{eq.22}) by definition are called entangled states~[38].
Thus the states are entangled if in formula (\ref{eq.22}) at least one
coefficient (or more) $p_{i}$ is negative which means that the
positive ones can take values greater than unity.

Let us discuss the condition for the system state to be separable.
According to the partial transpose criterion~[41],
the system is separable if the partial transpose of the matrix
$\rho_{AB}$ (\ref{eq.22}) gives a positive density matrix. This
condition is necessary but not sufficient. Let us discuss this
condition within the framework of positive-map matrix representation.
For example, for a spin-1/2 bipartite system, we have shown that the
map of a density matrix onto its transpose belongs to the matrix
semigroup of matrices ${\cal L}$. One should point out that this map
cannot be obtained by means of averaging with all positive probability
distributions $p_k$. On the other hand, it is obvious that the generic
criterion, which contains the Peres criterion as a partial case, can
be formulated as follows.

Let us map the density matrix $\rho_{AB}$ of a bipartite system onto
vector $\vec\rho_{AB}$. Let the vector $\vec\rho_{AB}$ be acted upon
by an arbitrary matrix, which represents the positive maps in
subsystems $A$ and $B$. Thus we get a new vector
\begin{equation}\label{K1}
\vec\rho_{AB}^{(p)}=\Big({\cal L}_A\otimes {\cal L}_B\Big) \vec\rho _{AB}.
\end{equation}
Let us construct the density matrix $\rho_{AB}^{(p)}$ using the
inverse map of the vectors onto matrices. If the initial density
matrix is separable, the new density matrix $\rho_{AB}^{(p)}$ must be
positive (and separable).

In the case of the bipartite spin-1/2 system, by choosing ${\cal
L}_A=1$ and with ${\cal L}_B$ being the matrix coinciding with the
matrix $g^{\alpha\beta}$, we obtain the Peres criterion as a partial
case of the criterion of separability formulated above. Thus, our
criterion means that the separable matrix keeps positivity under the
action of the tensor product of two semigroups. In the case of the
bipartite spin-1/2 system, the 16$\times$16 matrix of the semigroup
tensor product of positive contractive maps (\ref{R20}) is determined
by 24 parameters. Among these parameters, one can have some
correlations.

Let us discuss the positive map (\ref{R20}) which is determined by the
semigroup for the $n$-dimensional system. It can be realized also as
follows.

The $n$$\times$$n$ Hermitian generic matrix $\rho$ can be mapped onto
essentially real $n^2$-vector $\vec\rho$ by the map described above.
The complex vector $\vec\rho$ is mapped onto the real vector
$\vec\rho_{\rm r}$ by multiplying by the unitary matrix $S$, i.e.,
\begin{equation}\label{G1}
\vec\rho_{\rm r}=S\vec\rho,\qquad\vec\rho=S^{-1}\vec\rho_{\rm r}.
\end{equation}
The matrix $S$ is composed from $n$ unity blocks and the blocks
\begin{equation}\label{G2}
S_b^{(jk)}=\frac {1}{\sqrt 2} \left(
\begin{array}{cccc}
1&1 \\ -i&i\end{array} \right),  \end{equation} where $j$ corresponds
to a column and $k$ corresponds to a row in the matrix $\rho$.

For example, in the case $n=2$, one has the vector $\vec\rho_{\rm r}$
of the form
\begin{equation}\label{G3}
\vec\rho_{\rm r}= \left(
\begin{array}{c}
\rho_{11} \\
\sqrt 2\,\mbox{Re}\,\rho_{12}\\
\sqrt 2\,\mbox{Im}\,\rho_{12}\\
\rho_{22}
\end{array}
\right).  \end{equation} One has the equalities
\begin{equation}\label{G4}
\vec\rho_{\rm r}^2= \vec\rho^2=\mbox{Tr}\,\rho^2.
\end{equation}
The semigroup preserves the trace of the density matrix. Also the
discrete transforms, which are described by the matrix with diagonal
matrix blocks of the form
\begin{equation}\label{G5}
{\cal D}= \left(
\begin{array}{clcr}
1&0&0&0 \\
0&1&0&0 \\
0&0&-1&0 \\
0&0&0&1
\end{array}
\right),  \end{equation} preserve positivity of the density matrix.

For the spin case, the semigroup contains 12 parameters.

Thus, the direct product of the semigroup (\ref{R20}) and the discrete
group of the transform $D$ defines positive map preserving positivity
of the density operator. One can include also all the matrices which
correspond to other not completely positive maps. The considered
representation contains only real vectors and their real positive
transforms. This means that one can construct representation of
semigroup of positive maps by real matrices.

\section{Symbols, star-product and entanglement}

In this section, we describe how entangled states and separable states
can be studied using properties of symbols and density operators of
different kinds, e.g., from the viewpoint of the Wigner function or
tomogram. The general scheme  of constructing the operator symbols is
as follows~[36].

Given a Hilbert space $H$ and an operator $\hat A $ acting on this
space, let us suppose that we have a set of operators $\hat U({\bf
x})$ acting transitively on $H$ parametrized by $n$-dimensional
vectors ${\bf x}=(x_1,x_2,\ldots,x_n)$. We construct the $c$-number
function $f_{\hat A}({\bf x})$ (we call it the symbol of the operator
$\hat A$) using the definition
\begin{equation}\label{eq.1b}
f_{\hat A}({\bf x})=\mbox{Tr}\left[\hat A\hat U({\bf x})\right].
\end{equation}
Let us suppose that relation~(\ref{eq.1b}) has an inverse, i.e., there
exists a set of operators $\hat D({\bf x})$ acting on the Hilbert
space such that
\begin{equation}\label{eq.2b}
\hat A= \int f_{\hat A}({\bf x})\hat D({\bf x})~d{\bf x}, \qquad
\mbox{Tr}\, \hat A= \int f_{\hat A}({\bf x})\,\mbox{Tr}\, \hat D({\bf x})
~d{\bf x}.
\end{equation}
One needs a measure in $\bf x$ to define the integral in above
formulae. Then, we will consider relations~(\ref{eq.1b})
and~(\ref{eq.2b}) as relations determining the invertible map from the
operator $\hat A$ onto the function $f_{\hat A}({\bf x})$. Multiplying
both sides of Eq.~(\ref{eq.2}) by the operator $\hat U({\bf x}')$ and
taking the trace, one can satisfy the consistency condition for the
operators $\hat U({\bf x}')$ and $\hat D({\bf x})$
\begin{equation}\label{eq.2b'}
\mbox{Tr}\left[\hat U({\bf x}')\hat D({\bf x})\right]
=\delta\left({\bf x}'-{\bf x}\right).
\end{equation}
The consistency condition~(\ref{eq.2b'}) follows from the relation
\begin{equation}\label{eq.2aa}
f_{\hat A}({\bf x})=\int K({\bf x}, {\bf x}')f_{\hat A}({\bf x}')
\,d{\bf x}'.
\end{equation}
The kernel in~(\ref{eq.2aa}) is equal to the standard Dirac
delta-function, if the set of functions $f_{\hat A}({\bf x})$ is a
complete set.

In fact, we could consider relations of the form
\begin{equation}\label{eq.3b}
\hat A\rightarrow f_{\hat A}({\bf x})
\end{equation}
and
\begin{equation}\label{eq.4b}
f_{\hat A}({\bf x})\rightarrow\hat A.
\end{equation}
The most important property of the map is the existence of the
associative product (star-product) of the functions.

We introduce the product (star-product) of two functions $f_{\hat
A}({\bf x})$ and $f_{\hat B}({\bf x})$ corresponding to two operators
$\hat A$ and $\hat B$ by the relationships
\begin{equation}\label{eq.5b}
f_{\hat A\hat B}({\bf x})=f_{\hat A}({\bf x})* f_{\hat B} ({\bf
x}):=\mbox{Tr}\left[\hat A\hat B\hat U({\bf x}) \right].
\end{equation}
Since the standard product of operators on a Hilbert space is an
associative product, i.e., $\hat A(\hat B \hat C)=(\hat A\hat B)\hat
C$, it is obvious that formula~(\ref{eq.5b}) defines an associative
product for the functions $f_{\hat A}({\bf x})$, i.e.,
\begin{equation}\label{eq.6b}
f_{\hat A}({\bf x})*\Big(f_{\hat B}({\bf x}) *f_{\hat C}({\bf x})\Big)
= \Big(f_{\hat A}({\bf x})*f_{\hat B}({\bf x})\Big) *f_{\hat C}({\bf x}).
\end{equation}

Using formulae~(\ref{eq.1b}) and (\ref{eq.2b}), one can write down a
composition rule for two symbols $f_{\hat A}({\bf x})$ and $f_{\hat
B}({\bf x})$, which determines the star-product of these symbols. The
composition rule is described by the formula
\begin{equation}\label{eq.25b}
f_{\hat A}({\bf x})*f_{\hat B}({\bf x})= \int f_{\hat A}({\bf x}'')
f_{\hat B}({\bf x}')K({\bf x}'',{\bf x}',{\bf x})\,d{\bf x}'\,d{\bf
x}''.
\end{equation}
The kernel in the integral of (\ref{eq.25b}) is determined by the
trace of the product of the basic operators, which we use to construct
the map
\begin{equation}\label{eq.26b}
K({\bf x}'',{\bf x}',{\bf x})= \mbox{Tr}\left[\hat D({\bf x}'')
\hat D({\bf x}') \hat U({\bf x})\right].
\end{equation}
The kernel function satisfies the composition property $K*K=K$.

\section{Tomographic representation}

In this section, we will consider an example of the probability
representation of quantum mechanics~[42].
In the probability representation of quantum mechanics, the state is
described by a family of probabilities~[43--45].
According to the general scheme, one can introduce for the operator
$\hat A$ the function $f_{\hat A}({\bf x})$, where $${\bf
x}=(x_1,x_2,x_3)\equiv (X,\mu,\nu),$$ which we denote here as $w_{\hat
A}(X,\mu,\nu)$ depending on the position $X$ and the parameters $\mu$
and $\nu$ of the reference frame
\begin{equation}\label{eq.53}
w_{\hat A}(X,\mu,\nu)=\mbox{Tr}\left[\hat A \hat U({\bf x})\right].
\end{equation}
We call the function $w_{\hat A}(X,\mu,\nu)$ the tomographic symbol of
the operator $\hat A$. The operator $\hat U({\bf x})$ is given by
\begin{eqnarray}\label{eq.54}
 \hat U({\bf x})\equiv \hat U(X,\mu,\nu)&=&
\exp\left(\frac{i\lambda}{2}\left(\hat q\hat p +\hat p\hat q\right)\right)
\exp\left(\frac{i\theta}{2}\left(\hat q^2 +\hat p^2\right)\right)
\mid X\rangle\langle X\mid\nonumber\\ && \times\exp\left(-
\frac{i\theta}{2}\left(\hat q^2 +\hat p^2\right)\right)
\exp\left(-\frac{i\lambda}{2}\left(\hat q\hat p
+\hat p\hat q\right)\right)\nonumber\\&=&\hat
U_{\mu\nu}\mid X\rangle\langle X\mid\hat U_{\mu\nu}^\dagger.
\end{eqnarray}
The tomographic symbol is the homogeneous version of the Moyal
phase-space density. The angle $\theta$ and parameter $\lambda$ in
terms of the reference phase-space frame parameters are given by $$
\mu=e^{\lambda}\cos\theta, \qquad
\nu=e^{-\lambda}\sin\theta, $$ that is, $\hat q$ and $\hat p$ are
position and momentum operators
\begin{equation}\label{eq.54'}
\hat q\mid X\rangle=X\mid X\rangle
\end{equation}
and $\mid X\rangle\langle X\mid$ is the projection density. One has
the canonical transform of quadratures $$\hat X=\hat U_{\mu\nu}\,\hat
q\,\hat U^{\dagger}_{\mu\nu} =\mu \hat q+\nu\hat p,$$ $$ \hat P=\hat
U_{\mu\nu}\,\hat p\,\hat
U^{\dagger}_{\mu\nu}=\frac{1+\sqrt{1-4\mu^2\nu^2}}{2\mu}\,\hat p-
\frac{1-\sqrt{1-4\mu^2\nu^2}}{2\nu}\,\hat q. $$

Using the approach of [46]
one obtains the relationship
$$\hat U(X,\mu,\nu)=\delta(X-\mu \hat q-\nu\hat p).$$ In the case we
are considering, the inverse transform determining the operator in
terms of the tomogram [see Eq.~(\ref{eq.2b})] will be of the form
\begin{equation}\label{eq.55}
\hat A=\int w_{\hat A}(X,\mu,\nu) \hat D(X,\mu,\nu)\,dX\,d\mu\,d\nu,
\end{equation}
where
\begin{equation}\label{eq.56}
\hat D({\bf x})\equiv\hat D(X,\mu,\nu)=\frac{1}{2\pi}
\exp\left(iX-i\nu\hat p-i\mu\hat q\right).
\end{equation}

The trace of the above operator, which provides the kernel determining
the trace of an arbitrary operator in the tomographic representation,
reads
$$\mbox{Tr}\,\hat D({\bf x})= e^{iX}\delta (\mu)\delta(\nu).$$
The function $w_{\hat A}(X,\mu,\nu)$ satisfies the relation
\begin{equation}\label{eq.56'}
w_{\hat A}\left(\lambda X,\lambda \mu,\lambda\nu\right)
=\frac{1}{|\lambda|}\,w_{\hat A}(X,\mu,\nu).
\end{equation}
This means that the tomographic symbols of operators are homogeneous
functions of three variables.

If one takes two operators $\hat A_1$ and $\hat A_2$, which are
expressed through the corresponding functions by the formulas
\begin{eqnarray}
\hat A_1&=&\int w_{\hat A_1}(X',\mu',\nu')\hat
D(X',\mu',\nu')\,dX'\,d\mu' \,d\nu', \nonumber\\ &&\label{eq.57}\\
\hat A_2&=&\int w_{\hat A_2}(X'',\mu'',\nu'')
\hat D(X'',\mu'',\nu'')dX''\,d\mu''\,d\nu'',
\nonumber
\end{eqnarray}
and $\hat A$ denotes the product of $\hat A_1$ and $\hat A_2$, then
the function $w_{\hat A}(X,\mu,\nu)$, which corresponds to $\hat A$,
is the star-product of the functions $w_{\hat A_1}(X,\mu,\nu)$ and
$w_{\hat A_2}(X,\mu,\nu)$. Thus this product $$ w_{\hat
A}(X,\mu,\nu)=w_{\hat A_1}(X,\mu,\nu) *w_{\hat A_2}(X,\mu,\nu) $$
reads
\begin{equation}\label{eq.58}
w_{\hat A}(X,\mu,\nu)=\int w_{\hat A_1}({\bf x}'') w_{\hat A_2}({\bf
x}')K({\bf x}'',{\bf x}', {\bf x})\,d{\bf x''}\,d{\bf x'},
\end{equation}
with kernel given by
\begin{equation}\label{eq.59}
K({\bf x}'',{\bf x}',{\bf x})= \mbox{Tr}\left[\hat D(X'',\mu'',\nu'')
\hat D(X',\mu',\nu')\hat U(X,\mu,\nu)\right].
\end{equation}
The explicit form of the kernel reads
\begin{eqnarray}\label{KERNEL}
 &&K(X_1,\mu_1,\nu_1,X_2,\mu_2,\nu_2,X\mu,\nu)\nonumber\\
 &&=\frac{\delta\Big(\mu(\nu_1+\nu_2)-\nu(\mu_1+\mu_2)\Big)}{4\pi^2}
\,\exp\left(\frac{i}{2}\Big\{\left(\nu_1\mu_2-\nu_2\mu_1\right)
+2X_1+2X_2\right.\nonumber\\
&& \left.\left. -\left[\frac{1-\sqrt{1-4\mu^2\nu^2}}{\nu}
\left(\nu_1+\nu_2\right)+\frac{1+\sqrt{1-4\nu^2\mu^2}}{\mu}
\left(\mu_1+\mu_2\right)\right]X\right\}\right).\nonumber\\
&&
\end{eqnarray}

\section{Multipartite systems}

Let us assume that for multimode ($N$-mode) system one has
\begin{eqnarray}
\hat U(\vec y)=\prod_{k=1}^N\otimes\hat U\left(\vec x^{(k)}\right),
\label{P6}\\ \hat D(\vec
y)=\prod_{k=1}^N\otimes\hat D\left(\vec x^{(k)}\right),\label{P7}
\end{eqnarray}
where
\begin{equation}\label{P8}
\vec y=\Big(x_1^{(1)},x_2^{(1)},\ldots,x_m^{(1)},x_1^{(2)},x_2^{(2)},
\ldots,x_m^{(N)}\Big).
\end{equation}
This means that the symbol of the density operator of the composite
system reads
\begin{equation}\label{P9}
f_\rho(\vec y)=\mbox{Tr}\,\Big[\hat\rho\prod_{k=1}^N\otimes\hat U
(\vec x^{(k)})\Big].\end{equation} The inverse transform reads
\begin{equation}\label{P10}
\hat\rho=\int d\vec y\,f_\rho(\vec y)\prod_{k=1}^N\otimes
\hat D(\vec x^{(k)}), \qquad d\vec y=\prod_{k=1}^N\prod_{s=1}^mdx_s^{(k)}.
\end{equation}

Now we formulate the properties of the symbols in the case of
entangled and separable states, respectively.

Given a composite $m$-partite system with density operator $\hat\rho$.

If the nonnegative operator can be presented in the form of a
``probabilistic sum''
\begin{equation}\label{P14}
\hat\rho=\sum_{\vec z}{\cal P}(\vec z)\hat\rho_{\vec z}^{(a_1)}
\otimes \hat\rho_{\vec z}^{(a_2)}\otimes\cdots\otimes
\hat\rho_{\vec z}^{(a_m)},\end{equation} with positive
probability distribution function ${\cal P}(\vec z)$, where the
components of $\vec z$ can be either discrete or continuous, we call
the state a ``separable state.'' Without loss of generality, all
factors in the tensor products can be considered as projectors of rank
one. This means that the symbol of the state can be presented in the
form
\begin{equation}\label{P15}
f_\rho(\vec y)=\sum_{\vec z}{\cal P}(\vec
z)\prod_{k=1}^mf_\rho^{(a_k)} (\vec x_{k},\vec z).\end{equation}
Analogous formula can be written for the tomogram of separable state.
We point out that in the multipartite case one can introduce random
symbols and represent the symbol of separable density matrix of
composite system as mean value of pointwise products of symbols of
subsystem density operators. As in the bipartite case, one can use sum
over different indices but this sum can be always reduced to the sum
over only one index common for all the subsystems. It is important
that for separable state its symbol always can be represented as the
sum containing number of summants which is equal to dimensionality of
composite system. Each term in the sum is equal to mean value of
random projector. The random projector is constructed as the product
of diagonal projectors of rank one in each subsystem considered in
random local basis obtained by means of random unitary local
transforms.

\section{Spin tomography}

Below we concentrate on bipartite spin systems.

The tomographic probability (spin tomogram) completely determines the
density matrix of a spin state. It has been introduced in
[31,~32,~36].
The tomographic probability for the spin-$j$  state is defined via the
density matrix by the formula
\begin{equation}\label{eq.23c}
\langle jm\mid D^{\dagger }(g)\rho D(g)\mid jm\rangle=W^{(j)}(m,\vec{0}),
\qquad m=-j,-j+1,\ldots,j,
\end{equation} where $D(g)$ is the matrix of $SU(2)$-group
representation depending on the group element $g$ determined by three
Euler angles. It is useful to generalize the construction of spin
tomogram.

One can introduce unitary spin tomograms $w(m,u)$ by replacing in
above formula (\ref{eq.23c}) the matrix $D(g)$ by generic unitary
matrix $u$. For the case of higher spins $j=1,3/2,2,\ldots$, the
$n$${\times}$$n$ projector matrix
\begin{equation}\label{AA13}
\rho_1= \left(
\begin{array}{clcr}
1&0&\cdots&0\\0&0&\cdots&0\\
\cdot&\cdot&\cdots&\cdot\\
\cdot&\cdot&\cdots&\cdot\\
0&0&\cdots&0\end{array}\right),  \qquad n=2j+1
\end{equation}
has the unitary spin tomogram denoted as
\begin{equation}\label{AA14}
w_1(j,u)=|u_{11}|^2,\quad w_1(j-1,u)=|u_{12}|^2,\quad\ldots
\quad w_1(-j,u)=|u_{1n}|^2.\end{equation}
Other projectors
\begin{equation}\label{AA15}
\rho_k= \left(
\begin{array}{clcrc}
0&0&\cdots&\cdots&0\\
\cdot&\cdot&\cdots&\cdot&\cdot\\
0&\cdots&1&\cdots&0\\
\cdot&\cdot&\cdot&\cdots&\cdot\\
0&0&\cdot&\cdots&0\end{array}
\right),
\end{equation}
in which unity is located in $k$th column, have the tomogram
$w_k(m,u)$ of the form
\begin{equation}\label{AA16}
w_k(j,u)=|u_{k1}|^2,\quad w_k(j-1,u)=|u_{k2}|^2,\quad\ldots
\quad w_k(-j,u)=|u_{kn}|^2.\end{equation}
In connection with the decomposition of any density matrix in the form
\begin{equation}\label{AA17}
\rho=\sum_{jk}\rho_{jk}E_{jk},
\end{equation}
the unitary spin tomogram can be presented in form of the
decomposition
\begin{equation}\label{AA18}
w_\rho(m,u)=\sum_{jk}\rho_{jk}w_{jk}(m,u),
\end{equation}
where $w_{jk}(m,u)$ are basic unitary spin symbols of transition
operators $E_{jk}$ of the form
\begin{equation}\label{AA19}
w_{jk}(m,u)=\langle jm\mid u^\dagger E_{jk}u\mid jm\rangle.
\end{equation}
If one uses a map
\begin{equation}\label{AA20}
\rho\to\rho',
\end{equation}
the unitary spin tomogram is transformed as
\begin{equation}\label{AA21}
w_\rho(m,u)\to w'_\rho(m,u)=\sum_{jk}\rho'_{jk}w_{jk}(m,u).
\end{equation}
If the transform (\ref{AA20}) is a linear one
\begin{equation}\label{AA22}
\rho_{jk}\to\rho'_{jk}=L_{jk,ps}\rho_{ps},
\end{equation}
the transform reads
\begin{equation}\label{AA23}
w'_\rho(m,u)=\sum_{ps}\rho_{ps}w'_{ps}(m,u).
\end{equation}
Here
\begin{equation}\label{AA24}
w'_{ps}(m,u)=\sum_{jk}L_{jk,ps}w_{jk}(m,u)
\end{equation}
is the linear transform of the basic tomographic symbols    of the
operators $E_{jk}$.

Let us now discuss some properties of usual spin tomograms.

The set of the tomogram values for each $\vec 0$ is an overcomplete
set. We need only a finite number of independent locations which will
give information on the density matrix of the spin state. Due to the
structure of the formula, there are only two Euler angles involved.
They are combined into the unit vector
\begin{equation}\label{eq.24c} \vec{0}=(\cos \phi \sin
\vartheta ,\sin \phi \sin \vartheta ,\cos \vartheta ).\end{equation}
 This is the Hopf map from $S^3$ to $S^2$.

The physical meaning of the probability $W(m,\vec{0})$ is the
following.

It is the probability to find, in the state with the density matrix
$\rho $, the spin projection on direction  $\vec{0}$ equal to $m$. For
a bipartite system, the spin tomogram is defined as follows:
\begin{equation}\label{eq.25c}
W(m_{1}m_{2}\vec{0}_{1}\vec{0}_{2})=\langle j_{1}m_{1}j_{2}m_{2}
\mid D^{\dagger}(g_{1})D^{\dagger }(g_{2})\rho D(g_{1})D(g_{2})
\mid j_{1}m_{1}j_{2}m_{2}\rangle.
\end{equation} It completely determines the density matrix $\rho
$. It has the meaning of the joint probability distribution for spin
$j_{1}$ and $j_{2}$ projections $m_{1}$ and $m_{2}$  on directions $
\vec{0}_{1}$ and $\vec{0}_{2}.$ Since the map $\rho \rightleftharpoons
W$ is linear and invertible, the definition of separable system can be
rewritten in the following form for the decomposition of the joint
probability into a sum of products (of factorized probabilities):
\begin{equation}\label{eq.26c}
W(m_{1}m_{2}\vec{0}_{1}\vec{0}_{2})= \sum_{k}p_{k}W^{(k)}(m_{1}\vec{0}_{1})
\tilde{W}^{(k)}(m_{2}\vec{0}_{2}).\end{equation}
This form can be considered to formulate the criterion of separability
of the two-spin state.

One can present this formula in the form
\begin{equation}\label{eq.26cc}
W(m_{1}m_{2}\vec{0}_{1}\vec{0}_{2})=
\langle W(m_{1}\vec{0}_{1})
\tilde{W}(m_{2}\vec{0}_{2})\rangle,\end{equation}
where we interpret the positive numbers $p_k$ as probability
distributions. Thus separability means the possibility to represent
joint probability distribution in the form of average product of two
random probability distributions.

The state is separable iff the tomogram can be written in the form
(\ref{eq.26c}) with $\sum_{k}p_{k}=1$, $p_{k}\geq 0.$ It seems that we
simply use the definition but, in fact, we cast the problem of
separability into the form of the property of the positive joint
probability distribution of two random variables. This is an area of
probability theory and one can use the results and theorems on joint
probability distributions. If one does not use any theorem, one has to
study the solvability of relation (\ref{eq.26c}) considered as the
equation for unknown probability distribution $p_{k}$ and unknown
probability functions $W^{(k)}(m_{1}\vec{0}_{1})$ and
$W^{(k)}(m_{2}\vec{0}_{2})$.

\section{Example of spin-1/2 bipartite system}

For the spin-1/2 state, the generic density matrix can be presented in
the form
\begin{equation}\label{eq.27c}
\rho =\frac{1}{2}\left( 1+\vec{\sigma}\cdot \vec{n}\right), \qquad
\vec{n}=(n_{1},n_{2},n_{3}), \end{equation}
where $\vec{\sigma}$ are Pauli matrices and $\vec{n}^{2}\leq 1,$  with
the vector $\vec{n}$ for a pure state being the unit vector. This
decomposition means that we use as basis in 4-dimensional vector space
the vectors corresponding to the Pauli matrices and the unit matrix,
i.e.,
\begin{equation}\label{P1}
\vec\sigma_1= \left(
\begin{array}{c}
0 \\
1\\
1\\
0\end{array} \right),\qquad \vec\sigma_2= \left(
\begin{array}{c}
0 \\
-i\\
i\\
0\end{array} \right),\qquad \vec\sigma_3= \left(
\begin{array}{c}
1\\
0\\
0\\
-1\end{array} \right),\qquad \vec 1= \left(
\begin{array}{c}
1\\
0\\
0\\
1\end{array} \right).  \end{equation} The density matrix vector
\begin{equation}\label{P2}
\vec\rho= \left(
\begin{array}{c}
\rho_{11} \\
\rho_{12}\\
\rho_{21}\\
\rho_{22}\end{array} \right)\end{equation}
is decomposed in terms of the basis vectors
\begin{equation}\label{P3}
\vec\rho=\frac {1}{2}\,\Big(\vec 1+n_1\vec\sigma_1+n_2\vec\sigma_2
+n_3\vec\sigma_3\Big).
\end{equation}
This means that the spin tomogram of the spin-1/2 state can be given
in the form
\begin{equation}\label{eq.28c}
W\left(\frac{1}{2},\vec{0}\right)=\frac{1}{2} +\frac{\vec{n}\cdot
\vec{0}}{2}\,,
\qquad W\left(-\frac{1}{2},\vec{0}\right)=\frac{1}{2}
-\frac{\vec{n}\cdot\vec{0}}{2}\,. \end{equation}

We can consider tomograms of specific spin state. If the state is pure
state with density matrix
\begin{equation}\label{AA1}
\rho_+= \left(
\begin{array}{cl}
1&0\\0&0\end{array} \right),
\end{equation}
the spin tomogram $W(m,\vec 0)$, where $$m={\pm}\frac{1}{2},\qquad\vec
0=(\sin\theta\cos\varphi,\sin\theta\sin\varphi,\cos\theta)$$ has the
values
\begin{equation}\label{AA2}
W_+\left(\frac{1}{2},\vec 0\right)=\cos^2\frac{\theta}{2}\,,\quad
W_+\left(-\frac{1}{2},\vec 0\right)=\sin^2\frac{\theta}{2}\,.
\end{equation}
The tomogram of the pure state
\begin{equation}\label{AA3}
\rho_-= \left(
\begin{array}{clcr}
0&0\\0&1\end{array} \right),
\end{equation}
has the values
\begin{eqnarray}\label{AA4}
W_-\left(\frac{1}{2},\vec 0\right)&=&\sin^2\frac{\theta}{2}=
\cos^2\frac{\pi-\theta}{2}\,,\nonumber\\
&&\\ W_-\left(-\frac{1}{2},\vec 0\right)&=&\cos^2\frac{\theta}{2}
=\sin^2\frac{\pi-\theta}{2}\,.\nonumber
\end{eqnarray}
The spin tomogram of the diagonal density matrix
\begin{equation}\label{AA5}
\rho_d= \left(
\begin{array}{clcr}
\rho_{11}&0\\0&\rho_{22}\end{array} \right)
\end{equation}
equals
\begin{equation}\label{AA6}
W_d(m,\vec 0)=\rho_{11}W_+(m,\vec 0_+)+\rho_{22}W_-(m,\vec 0_-).
\end{equation}
The generic density matrix which has eigenvalues $\rho_{11}$ and
$\rho_{22}$ can be presented in the form
\begin{equation}\label{AA7}
u_0\rho_du^\dagger_0,
\end{equation}
where the unitary matrix $u_0$ has columns containing components of
normalized eigenvectors of the density matrix $\rho$.

This means that the tomogram of the state with the matrix $\rho$ reads
\begin{equation}\label{AA8}
W_\rho(m,\vec{0})=\langle m\mid u^\dagger u_0\rho_du_0^\dagger u\mid
m\rangle .\end{equation} The elements of the group can be combined
\begin{equation}\label{AA9}
u^\dagger u_0=\tilde u.
\end{equation}
Thus the tomogram becomes
\begin{equation}\label{AA10}
W_\rho(m,\vec 0)=W_d(m,\vec 0'),
\end{equation}
where the angle $\vec 0'$ corresponds to the Euler angle calculated
from the product of two unitary matrices $u_0^\dagger u$.

One can use the property of numbers  $\rho_{11}$ and $\rho_{22}$ to
interpret formula (\ref{AA6}) as averaging
\begin{equation}\label{AA11}
W_d(m,\vec 0)=\langle W(m,\vec 0')\rangle,
\end{equation}
where one interprets two functions $W_+(m,\vec 0)$ and $W_-(m,\vec
0')$ as the realization of ``random'' probability distribution
function $W_\pm(m,\vec 0)$. Then one has
\begin{equation}\label{AA12}
W_\rho(m,\vec 0)=\langle W(m,\vec 0')\rangle.
\end{equation}
The new vector $\vec 0'$ has the parameter $\theta'$ obtained from the
initial parameter $\theta$ by action of the unitary matrix on the
initial unitary matrix $u$.

Inserting these probability values into relation (\ref{eq.26c}) for
each value of $k$, we get the relationships
\begin{eqnarray}\label{eq.29c}
W\left(\frac{1}{2},\frac{1}{2},\vec{0}_{1},\vec{0}_{2}\right)&
=&\frac{1}{4}+\frac{1}{2}\left(
\sum_{k}p_{k}\vec{n}_{k}\right) \cdot \vec{0}_{1}+\frac{1}{2}\left(
\sum_{k}p_{k}\vec{n}_{k}^*\right) \cdot \vec{0}_{2}\nonumber\\
&&+\sum_{k}p_{k}\left( \vec{n}_{k}\cdot \vec{0}_{1}\right) \left(
\vec{n}_{k}^*\cdot \vec{0}_{2}\right),
\end{eqnarray}
\begin{eqnarray}\label{eq.30c}
W\left(\frac{1}{2},-\frac{1}{2},\vec{0}_{1},\vec{0}_{2}\right)&
=&\frac{1}{4}+\frac{1}{2}\left(\sum_{k}p_{k}\vec{n}_{k}\right)
 \cdot \vec{0}_{1}-\frac{1}{2}\left(\sum_{k}p_{k}\vec{n}_{k}^*\right)
 \cdot \vec{0}_{2}\nonumber\\ &&-\sum_{k}p_{k}\left( \vec{n}_{k}\cdot
\vec{0}_{1}\right) \left( \vec{n}_{k}^*\cdot \vec{0}_{2}\right),
\end{eqnarray}
\begin{eqnarray}\label{eq.31c}
W\left(-\frac{1}{2},\frac{1}{2},\vec{0}_{1},\vec{0}_{2}\right)&
=&\frac{1}{4}-\frac{1}{2}\left(\sum_{k}p_{k}\vec{n}_{k}\right)
\cdot \vec{0}_{1}+\frac{1}{2}\left(\sum_{k}p_{k}\vec{n}_{k}^*\right)
\cdot \vec{0}_{2}\nonumber\\&&-\sum_{k}p_{k} \left( \vec{n}_{k}\cdot
\vec{0}_{1}\right) \left( \vec{n}_{k}^*\cdot \vec{0}_{2}\right).
\end{eqnarray}
One has the normalization property
\begin{equation}\label{eq.32c}
\sum_{m_{1},\,m_{2}=-{1}/{2}}^{{1}/{2}}W(m_{1}m_{2}
\vec{0}_{1}\vec{0}_{2})=1.\end{equation}
One easily gets
\begin{equation}\label{eq.33c}
W\left(\frac{1}{2},\frac{1}{2},\vec{0}_{1},\vec{0}_{2}\right)
+W\left(\frac{1}{2},-\frac{1}{2},\vec{0}_{1},\vec{0}_{2}\right)=
\frac{1}{2}+\left(\sum_{k}p_{k}\vec{n}_{k}\right) \cdot \vec{0}_{1}.\\
\end{equation} This means that the
derivative in $\vec{0}_{1}$ on the left-hand side gives
\begin{equation}\label{eq.34c}
\frac{\partial }{\partial \vec{0}_{1}}\left[ W\left(\frac{1}{2},
\frac{1}{2},\vec{0}_{1},\vec{0}_{2}\right)+W\left(\frac{1}{2},
-\frac{1}{2},\vec{0}_{1},\vec{0}_{2}\right)\right]
=\left( \sum_{k}p_{k}\vec{n}_{k}\right).\\\end{equation}
Analogously
\begin{equation}\label{eq.35c}
\frac{\partial }{\partial \vec{0}_{2}}\left[ W\left(\frac{1}{2},
\frac{1}{2},\vec{0}_{1},\vec{0}_{2}\right)+W\left(-\frac{1}{2},
\frac{1}{2},\vec{0}_{1},\vec{0}_{2}\right)\right] =\left(
\sum_{k}p_{k}\vec{n}_{k}^{\left( \star \right) }\right).\\\
\end{equation}
Taking the sum of (\ref{eq.30c}) and (\ref{eq.31c})) one sees that
\begin{eqnarray}\label{eq.36c}
&&\frac{1}{2} \frac{\partial }{\partial \vec{0}_{i}}\frac{\partial
}{\partial \vec{0}_{j}}\left[
W\left(\frac{1}{2},-\frac{1}{2},\vec{0}_{1},\vec{0}_{2}\right)
+W\left(-\frac{1}{2},\frac{1}{2},\vec{0}_{1},\vec{0}_{2}\right)\right]
\nonumber\\
&&=-\sum_{k}p_{k}(n_{k})_{i}(n_{k}^{\left( \star \right) })_{j}.
\end{eqnarray}

Since we look for the solution where $p_{k}\geq 0$, we can introduce
\begin{equation}\label{eq.37c}
\vec{N}_{k}=\sqrt{p_{k}}\vec{n}_{k}, \qquad \vec{N}_{k}^{\left(
\star \right)}=\sqrt{p_{k}}\vec{n}_{k}^{\left( \star \right) }.
\end{equation}
This means that the derivative in (\ref{eq.36c}) can be presented as a
tensor
\begin{equation}\label{eq.38c}
-T_{ij}=\sum_{k}(N_{k})_{i}(N_{k}^{\left(\star \right) })_j.
\end{equation}
One has \begin{equation}\label{eq.39c}
\sum_{k}p_{k}\vec{n}_{k}^{{}}=\sum_{k}\sqrt{p_{k}}\vec{N}_{k},
\end{equation}
\begin{equation}\label{eq.40c}
\sum_{k}p_{k}\vec{n}_{k}^{\star }=\sum_{k}\sqrt{p_{k}}
\vec{N}_{k}^{\left( \star \right) }.\end{equation} The conditions
of solvability of the obtained equations is a criterion for
separability or entanglement of a bipartite quantum spin state. Using
the arguments on the representation of the tomogram (tomographic
symbol) as sum of random basic projector symbols we get that for two
qubits the separable state has the tomogram with following properties.
All four values of joint probability distribution function are equal
to mean values of product of two cosine of two different angles
squared, product of sine of two different angles squared and product
of sine and cosine squared, respectively. The entangled matrix does
not provide such structure.

As an example, we consider the Werner state. For the Werner state
(see, e.g., [47])
with the density matrix
\begin{eqnarray}\label{eq.41c}
&&\rho _{AB}=\left(
\begin{array}{cccc}
\frac{1+p}{4} & 0 & 0 & \frac{p}{2} \\ 0 & \frac{1-p}{4} & 0 & 0 \\
 0 & 0 &\frac{1-p}{4} & 0 \\ \frac{p}{2} & 0 & 0 & \frac{1+p}{4}
\end{array}
\right),\nonumber\\
&&\\ &&\rho _{A}=\rho _{B}=\frac{1}{2}\left(
\begin{array}{cc}
1 & 0 \\ 0 & 1\nonumber
\end{array}
\right), \end{eqnarray} one can reconstruct the known results
that for $p<{1}/{3}$ the state is separable and for $p>{1}/{3}$ the
state is entangled, since in the decomposition of the density operator
in the form (\ref{eq.26c}) the state
\begin{equation}\label{eq.42c}
\rho _{0}=\frac{1}{4}\left(
\begin{array}{cccc}
1 & 0 & 0 & 0 \\ 0 & 1 & 0 & 0 \\ 0 & 0 & 1 & 0 \\
0 & 0 & 0 & 1\end{array} \right)
\end{equation} has the weight $p_{0}=(1-3p)/{4}$.

For $p>{1}/{3}$, the coefficient $p_{o}$ becomes negative.

There is some extension of the presented consideration.

Let us consider the state with the density matrix (nonnegative and
Hermitian)
\begin{equation}\label{W1}
\rho =\left(
\begin{array}{cccc}
R_{11} & 0 & 0 & R_{12} \\ 0 & \rho_{11} & \rho_{12} & 0 \\
0 & \rho_{21} & \rho_{22} & 0 \\
R_{21} & 0 & 0 & R_{22}\end{array} \right),
\qquad \mbox{Tr}\,\rho=1. \end{equation} Using
the procedure of mapping the matrix onto vector $\vec\rho$ and
applying to the vector the nonlocal linear transform corresponding to
the Peres partial transpose and making the inverse map of the
transformed vector onto the matrix, we obtain
\begin{equation}\label{W2}
\rho^m =\left(
\begin{array}{cccc}
R_{11} & 0 & 0 & \rho_{12} \\ 0 & \rho_{11} & R_{12} & 0 \\
0 & R_{21} & \rho_{22} & 0 \\
\rho_{21} & 0 & 0 & R_{22}\end{array} \right). \end{equation}
In the case of separable matrix $\rho$, the matrix $\rho^m$ is a
nonnegative matrix. Calculating the eigenvalues of $\rho^m$ and
applying the condition of their positivity, we get
\begin{equation}\label{W3}
R_{11}R_{22}\geq |\rho_{12}|^2,\qquad
\rho_{11}\rho_{22} \geq |R_{12}|^2.
\end{equation}
Violation of these inequalities gives a signal that $\rho$ is
entangled. For Werner state (\ref{eq.41c}), Eq.~(\ref{W3}) means
\begin{equation}\label{W4}
1+p>0,\qquad 1-p>2p,
\end{equation}
which recovers the condition of separability $p<1/3$ mentioned above.

The joint probability distribution (\ref{eq.25c}) of separable state
is positive after making the local and nonlocal (partial
transpose-like) transforms connected with positive map semigroup. But
for entangled state, function (\ref{eq.25c}) can take negative values
after making this map in the function and replacing on the right-hand
side of this equality the product of two matrices $D(g)$ by generic
unitary transform $u$. This is a criterion of entanglement in terms of
unitary spin tomogram of the state of multiparticle system.

A simpler and more transparent case is the generalized Werner
model with density matrix
\begin{equation}\label{30}
\rho=\frac{1}{4}\left(1+\mu_1\sigma_1\otimes
\tau_1+\mu_2\sigma_2\otimes\tau_2+
\mu_3\sigma_3\otimes\tau_3\right).
\end{equation}
Here the density matrix is expressed in terms of tensor products of
two sets of Pauli matrices $\sigma_k$ and $\tau_k$ $(k=1,2,3)$, which
are chosen in the standard form.

Its eigenvalues are $$1-\mu_1-\mu_2-\mu_3,\quad
1+\mu_1+\mu_2-\mu_3,\quad 1+\mu_1-\mu_2+\mu_3,\quad
1-\mu_1+\mu_2+\mu_3.$$ These eigenvalues are related to the vertices
of a regular tetrahedron. The partially time-reversed density matrix
is
\begin{equation}\label{31}
\widetilde\rho=\frac{1}{4}\left(1-\mu_1\sigma_1\otimes\tau_1-
\mu_2\sigma_2\otimes\tau_2-\mu_3\sigma_3\otimes\tau_3\right),
\end{equation}
which may be viewed as
\begin{equation}\label{32}
\widetilde\rho=L^{(1)}\otimes L^{(2)}\rho\qquad\mbox{with}\qquad
L^{(1)}\rho^{(1)}=\rho^{(1)},\quad L^{(2)}\rho^{(2)}=1-\rho^{(2)}.
\end{equation}
The eigenvalues of this are $$1+\mu_1+\mu_2+\mu_3,\quad
1+\mu_1-\mu_2-\mu_3,\quad 1-\mu_1+\mu_2-\mu_3,\quad
1-\mu_1-\mu_2+\mu_3.$$ These form an inverted tetrahedron and they
have the common domain which is a regular octahedron. The unitary spin
tomograms can be written down by inspection and we may verify that all
the relations required by the separability criterion (see the next
section for details) are satisfied by any point inside the octahedron
for $\rho$ and for $L^{(1)}$$\otimes$$L^{(2)}\rho$ but the relations
connected with positivity condition expressed in terms of positivity
of unitary spin tomogram fail when it lies outside.

\section{Tomogram of the group $U(n)$}

In this section we discuss in more detail the separability criterion
using introduced notion of unitary spin tomogram.

In order to formulate a criterion of separability for a bipartite spin
system with spin $j_1$ and $j_2$, we introduce the tomogram $w(\vec l,
\vec m,g^{(n)})$ for the group $U(n)$, where
$$n=n_1n_2,\qquad n_1=2j_1+1,\qquad n_2=2j_2+1,$$ and $g^{(n)}$
are parameters of the group element. Vectors $\vec l$ and $\vec m$
label a basis $\mid \vec l,\vec m\rangle$ of the fundamental
representation of the group $U(n)$. For example, since this
representation is irreducible, being reduced to the representation
of the $U(n_1)\otimes U(n_2)$ subgroup of the group $U(n)$, the
basis can be chosen as the product of basis vectors:
\begin{equation}\label{U1}
\mid j_1,m_1\rangle \mid j_2,m_2\rangle =\mid j_1,j_2,m_1,m_2\rangle.
\end{equation}
Due to the irreducibility of this representation of the group $U(n)$
and its subgroup, there exists a unitary transform
$u_{j_1j_2m_1m_2}^{\vec l\vec m}\mid \vec l,\vec m\rangle$ such that
\begin{eqnarray}
&&\mid j_1,j_2,m_1,m_2\rangle=\sum_{\vec l\vec m}
u_{j_1j_2m_1m_2}^{\vec l\vec m}\mid \vec
l,\vec m\rangle,\label{U2}\\ &&\mid \vec l\vec m \rangle
=\sum_{m_1m_2}(u^{-1})^{\vec l\vec
m}_{j_1j_2m_1m_2}\mid j_l,j_2,m_1,m_2\rangle. \label{U3}\end{eqnarray}
One can define the $U(n)$ tomogram for a Hermitian nonnegative
$n$$\times$$n$ density matrix $\rho$, which belongs to the Lie algebra
of the group $U(n)$, by a generic formula
\begin{equation}\label{U4}
w(\vec l,\vec m,g^{(n)})=\langle\vec l,\vec m\mid U^\dagger(g^{(n)})
\rho U(g^{(n)})\mid\vec l,\vec m\rangle.
\end{equation}
Formula~(\ref{U4}) defines the tomogram in the basis $\mid\vec l,\vec
m\rangle$ for arbitrary irreducible representation of the unitary
group. But below we focus only on tomograms connected with spins.

Let us define the $U(n)$ tomogram using the basis $\mid
j_1,j_2,m_1,m_2\rangle$ namely for fundamental representation, i.e.,
\begin{eqnarray}\label{U5}
&&w^{(j_1,j_2)}(m_1, m_2,g^{(n)})\nonumber\\ &&=\langle
j_1,j_2,m_1,m_2
\mid U^\dagger(g^{(n)})\rho U(g^{(n)})\mid j_1,j_2,m_1,m_2\rangle.
\end{eqnarray}
This unitary spin tomogram becomes the spin-tomogram~[34]
for the $g^{(n)}\in U(2)\otimes U(2)$ subgroup of the group $U(n)$.
The properties of this tomogram follow from its definition as the
joint probability distribution of two random spin projections
$m_1,m_2$ depending on $g^{(n)}$ parameters.

One has the normalization condition
\begin{equation}\label{U6}
\sum_{m_1,m_2}w^{(j_1,j_2)}(m_1, m_2,g^{(n)})=1.
\end{equation}
Also all the probabilities are nonnegative, i.e.,
\begin{equation}\label{U7}
w^{(j_1,j_2)}(m_l, m_2,g^{(n)})\geq 0.
\end{equation}
Due to this, one has
\begin{equation}\label{U8}
\sum_{m_1,m_2}|w^{(j_1,j_2)}(m_l, m_2,g^{(n)})|=1.
\end{equation}
For the spin-tomogram,
\begin{equation}\label{U9}
g^{(n)}\rightarrow \Big(\vec O_1,\vec O_2\Big)
\end{equation}
and
\begin{equation}\label{U10}
w^{(j_1,j_2)}(m_l, m_2,g^{(n)})\rightarrow w(m_1,m_2,\vec O_1,\vec O_2).
\end{equation}

The separability and entanglement condition discussed in the previous
section for a bipartite spin-tomogram can be considered also from the
viewpoint of the properties of a $U(n)$ tomogram. If the two-spin
$n$$\times$$n$ density matrix $\rho$ is separable, it remains
separable under the action of the generic positive map of the
subsystem density matrices. This map can be described as follows.

Let $\rho$ be mapped onto vector $\vec \rho$ with $n^2$ components.
The components are simply ordered rows of the matrix $\rho$, i.e.,
\begin{equation}\label{U11}
\vec\rho=\Big(\rho_{11},\rho_{12},\ldots,\rho_{1n},\rho_{21},\rho_{22},
\ldots,\rho_{nn},\Big).
\end{equation}
Let the $n^2$$\times$$n^2$ matrix $L$ be taken in the form
\begin{equation}\label{U12} L=\sum_sp_sL_s^{(j_1)}\otimes
L_s^{(j_2)},\qquad p_s\geq 0,\quad \sum_s p_s=1,
\end{equation}
where the $n_1$$\times$$n_1$ matrix $L_s^{(j_1)}$ and the
$n_2$$\times$$n_2$ matrix $L_s^{(j_2)}$ describe the positive maps of
density matrices of spin-$j_1$ and spin-$j_2$ subsystems,
respectively. We map vector $\vec\rho$ onto vector $\vec\rho_L$
\begin{equation}\label{U13}
\vec\rho_L=L\vec\rho
\end{equation}
and construct the $n$$\times$$n$ matrix $\rho_L$, which corresponds to
the vector $\vec\rho_L$. Then we consider the $U(n)$ tomogram of the
matrix $\rho_L$, i.e.,
\begin{eqnarray}\label{U14}
&&w^{(j_1,j_2)}_L(m_l, m_2,g^{(n)})\nonumber\\ &&=\langle
j_1,j_2,m_1,m_2\mid U^\dagger(g^{(n)})\rho_L U(g^{(n)})\mid
j_1,j_2,m_l,m_2\rangle.
\end{eqnarray}
Using this tomogram we introduce the function
\begin{equation}\label{U15}
F(g^{(n)},L)=\sum_{m_1,m_2}\left|w_L^{(j_1,j_2)}(m_1,m_2,g^{(n)})\right|.
\end{equation} For separable states, this function does not depend
on the $U(n)$-group parameter $g^{(n)}$ and positive-map matrix
elements of the matrix $L$.

For the normalized density matrix $\rho$ of the bipartite spin-system,
this function reads
\begin{equation}\label{U16}
F(g^{(n)},L)=1.
\end{equation}
For entangled states, this function depends on $g^{(n)}$ and $L$ and
is not equal to unity. This property can be chosen as a necessary and
sufficient condition for separability of bipartite spin-states. We
introduce also tomographic purity parameter $\mu_k$ of $k$th order by
the formula
$$\mu_k(g^{(n)},L)=\sum_{m_1m_2}\left|w_L^{(j_1,j_2)}
(m_1,m_2,g^{(h)})\right|^k.$$ For identity semigroup element $L$ and
specific $g_0^{(n)}$ unitary transform diagonalizing the density
matrix, the tomographic purity $\mu_2$ is identical to purity
parameter of the state $\rho$. The parameters for $k=2,3,\ldots$,
correspond to Tr$\,\rho^{k+1}$.

In fact, the formulated approach can be extended to multipartite
systems too. The generalization is as follows.

Given $N$ spin-systems with spins $j_1,j_2,\ldots,j_N$, let us consider
the group $U(n)$ with
\begin{equation}\label{U17}
n=\prod_{k=1}^Nn_k,\qquad n_k=2j_k+1.
\end{equation}
Let us introduce the basis
\begin{equation}\label{U18}
\mid \vec m\rangle=\prod_{k=1}^N\mid j_km_k\rangle
\end{equation}
in the linear space of the fundamental representation of the group
$U(n)$. We define now the $U(n)$ tomogram of a state with the
$n$$\times$$n$ matrix $\rho$:
\begin{equation}\label{U19}
w_\rho(\vec m,g^{(n)})=\langle \vec m\mid U^\dagger(g^{(n)})
\rho U(g^{(n)})\mid\vec m\rangle.
\end{equation}
For a positive Hermitian matrix $\rho$ with $\mbox{Tr}\,\rho=1$, we
formulate the criterion of separability as follows.

Let the map matrix $L$ be of the form
\begin{equation}\label{U20}
L=\sum_sp_s\Big(\prod_{k=1}^N\otimes L_s^{(k)}\Big), \qquad
p_s\geq 0,\quad \sum_s p_s=1,
\end{equation}
where $L_s^{(k)}$ is the positive-map matrix of the density matrix of
the $k$th spin subsystem. We construct the matrix $\rho_L$ as in the
case of the bipartite system using the matrix $L$. The function
\begin{equation}\label{U21}
F(g^{(n)},L)=\sum_{\vec m}|w_{\rho_L}(\vec m,g^{(n)})|\geq 1
\end{equation}
is equal to unity for separable state and depends on the matrix $L$
and $U(n)$-parameters $g^{(n)}$ for entangled states.

This criterion can be applied also in the case of continuous
variables, e.g., for Gaussian states of photons. Function (\ref{U21})
can provide the measure of entanglement. Thus one can use the maximum
value (or a mean value) of this function as a characteristic of
entanglement. In the previous section, we considered the generalized
Werner states. Using the above criterion, one can get the domain of
values of the parameters of the states for which one has separability
or entanglement. In fact, the separability criterion is related to the
following positivity criterion of finite or infinite (trace class)
matrix $A$. The matrix $A$ is positive iff the sum of moduli of
diagonal matrix elements of the matrix $UAU^\dagger$ is equal to a
positive trace of the matrix $A$ for an arbitrary unitary matrix $U$.

\section{Dynamical map and purification}

In this section, we consider the connection of positive maps with
purification procedure. In fact, formula
\begin{equation}\label{Pu1}
\rho\to\rho'=\sum_kp_kU_k\rho U_k^\dagger,\end{equation} where
$U_k$ are unitary operators, can be considered in the form
\begin{equation}\label{Pu2}
\rho\to\rho'=\sum_kp_k\rho_k,\qquad p_k\geq 0,\qquad
\sum_kp_k=1.\end{equation} Here the density operators $\rho_k$
read
\begin{equation}\label{Pu3}\rho_k=U_k\rho U_k^\dagger,\end{equation}
and the maps which are not sufficiently general keep the most
degenerate density matrix fixed. This form is the form of
probabilistic addition. This mixture of density operators can be
purified with the help of a fiducial rank one projector $P_0$
\begin{equation}\label{Pu4}\rho'\to\rho''=N\left[\sum_{kj}
\sqrt{p_kp_j}\,\frac{\rho_kP_0\rho_j}{\sqrt{
\mbox{Tr}\,\rho_kP_0\rho_jP_0}}\right],\end{equation}
where $N$ is a normalization constant
\begin{equation}\label{Pu5}
N^{-1}=\mbox{Tr}\,\left(\sum_{kj}\sqrt{p_kp_j}\,
\frac{\rho_kP_0\rho_j}{\sqrt{\mbox{Tr}\,\rho_kP_0\rho_jP_0}}
\right).\end{equation}
The normalization is unnecessary if all $\rho_k$ are mutually
orthogonal. We call this map a purification map. It maps the density
matrix of mixed state on the density matrix of pure state.

The map (\ref{Pu1}) could be interpreted as the evolution in time of
the initial matrix $\rho_0$ considering unitary operators $U_k(t)$
depending on time. Thus one has
\begin{equation}\label{Pu6}
\rho_0\to\rho(t)=\sum_kp_kU_k(t)\rho_0U_k^\dagger(t).\end{equation}
In this case, the purification procedure provides the dynamical
map of a pure state
\begin{equation}\label{Pu7}
\mid\psi_0\rangle\langle\psi_0\mid\to\mid \psi(t)\rangle\langle
\psi(t)\mid,\end{equation} where $\mid \psi(t)\rangle$ obeys a
nonlinear equation and, in the general case, this map does not define
a one parameter group of transformations not even locally.

For some specific cases, the evolution (\ref{Pu6}) can be described by
a semigroup. The density matrix (\ref{Pu6}) obeys then a first-order
differential equation in time for this case~[27--29].

More specifically, the reason why there is no differential equation in
time for the generic case is due to the absence of the property
\begin{equation}\label{Pu8}
\rho_{ij}(t_2)=\sum_{mn}K_{ij}^{mn}(t_2,t_1)\rho_{mn}(t_1),
\end{equation}
where the kernel of evolution operator satisfies
\begin{equation}\label{Pu9}
K_{ij}^{mn}(t_3,t_2)K_{mn}^{pq}(t_2,t_1)=K_{ij}^{pg}(t_3,t_1).
\end{equation}

Thus, via a purification procedure and a dynamical map applied to a
density matrix we get a pure state (nonlinear dynamical map). This map
can be used in nonlinear models of quantum evolution. Many linear
positive maps both completely positive and not completely positive are
contractive. We define a positive map $L$ as ``contractive'' or
``dilating'' if $\mbox{Tr}\,(L\rho)^2\leq\mbox{Tr}\,(\rho)^2$ or
$\mbox{Tr}\,(L\rho)^2\geq\mbox{Tr}\,(\rho)^2$, respectively. This
means, for example, that purity parameter $\mu=\mbox{Tr}\,\rho^2$
after performing the positive map generically becomes smaller. There
are maps for which the purity parameter is preserved, for example,
\begin{equation}\label{AA25}
\rho\to\rho^{\rm tr},\qquad \rho\to -\rho +\frac{2}{N}\,1.
\end{equation}
These linear maps include also unitary transform
\begin{equation}\label{AA26}
\rho\to u\rho u^\dagger.
\end{equation}
There are maps  which provide dilation. For qubit system, matrix
\begin{equation}\label{HHH15}
L=\left(
\begin{array}{cccc}
1& 0 & 0 & 1 \\ 0 & 0 & 0 & 0 \\ 0 & 0&0 & 0 \\ 0 & 0 & 0 &
0\end{array}
\right).
\end{equation}
acting on arbitrary vector $\vec\rho_0$ corresponding to a density
matrix $\rho_0$ gives the pure state
\begin{equation}\label{HHHH15}
\rho_f =\left(
\begin{array}{cc}
1& 0 \\ 0 & 0\end{array}
\right).
\end{equation}
The matrices $L_\varepsilon$ the inverse matrices exist for
$\varepsilon\neq  0$. But these inverse matrices do not provide
positive trace preserving maps. Since the purification procedure
provides a positive map, which increases the purity parameter, the
composition of linear map with the purification map provides the
possibility to recover the initial density matrix $\rho$ which was the
object of action of a positive linear map. It means that the
purification map $\hat L_p$ can give
\begin{equation}\label{AA28}
\hat L_{P_0}(L\rho)=1\rho
\end{equation}
for any density matrix $\rho$ but the choice of fiducial projector
depends on $\rho$ (the initial condition).

Thus one has also for completely positive maps
\begin{equation}\label{AA29}
\rho\to\rho'=\sum_k\rho_k',\qquad
\rho_k=V_k\rho V_k^\dagger,\qquad
\sum_kV_k^\dagger V_k=1.
\end{equation}
Making polar decomposition
$$\rho_k=\sqrt{\rho_{0k}}U_k,\qquad U_kU_k^\dagger=1,
\qquad \rho_{0k}\geq 0$$
and introducing the positive numbers $p_k=\mbox{Tr}\,\rho_{0k}$, we
construct the map
\begin{equation}\label{AA30}
\rho'\to\rho''=\left\{\sum_{kj}\sqrt{p_kp_j}
\frac{\tilde\rho_kP_0\tilde\rho_j+\tilde\rho_jP_0\tilde\rho_k}
{\sqrt{T\tilde\rho_kP_0\tilde\rho_jP_0}}\right\},
\quad \sum_kp_k=1, \quad p_k\tilde\rho_k=\rho_k.\end{equation}
The matrix $\rho''$ is a matrix of rank one for any rank one fiducial
projector $P_0$. The projector $P_0$ is restricted to be not
orthogonal to the generic matrix $\rho_k$. Taking $N$ orthogonal
projectors $P_0^{(s)}~(s=1,2,\ldots,N)$ and obtaining $N$ projectors
$\rho''_s$, one can combine them in order to get the initial matrix
$\rho$. It means that one can take convex sum of the $N$ pure states
$\rho''_s$ to recover the initial mixed state $\rho$. Another way to
make the state with higher purity was demonstrated using the modified
purification procedure in [48].
For qubit state, one has
\begin{equation}\label{AA31}
\rho=p_1\rho_1+p_2\rho_2+\kappa\sqrt{p_1p_2}\,\frac{\rho_1P_0\rho_2+
\rho_2P_0\rho_1}{\sqrt{\mbox{Tr}\,\rho_1P_0\rho_2P_0}}\,,
\quad p_1+p_2=1,
\end{equation}
where the decoherence parameter $0\leq\kappa\leq 1$ is used. If
$\kappa\sim 1$, we increase purity.

Let us discuss the map (\ref{AA29}) using its matrix form, i.e.,
\begin{equation}\label{HH1}
\rho_{\alpha\beta}\to\rho'_{\alpha\beta}=\sum_{ij}
{\cal L}_{\alpha\beta,ij}\rho_{ij}.
\end{equation}
The matrix ${\cal L}_{\alpha\beta,ij}$ is expressed in terms of the
matrices $V_k$ as
\begin{equation}\label{HH2}
{\cal L}_{\alpha\beta,ij}=\sum_k(V_k)_{\alpha i}(V_k^*)_{\beta j}.
\end{equation}
One can construct another positive map~[49]
\begin{equation}\label{HH3}
\rho\to\rho'=\sum_kr_k\mbox{Tr}\left(R_k\rho\right),
\end{equation}
where $r_k$ are density matrices and $R_k$ are positive operators
satisfying the normalization condition
\begin{equation}\label{HH4}
\sum_kR_k=1.
\end{equation}
The matrix corresponding to this map (called entanglement breaking
map~[50])
reads
\begin{equation}\label{HH5}
{\cal L}^b_{\alpha\beta,ij}=\sum_k(r_k)_{\alpha\beta}(R_k^*)_{ij}.
\end{equation}
The entanglement breaking map is contractive positive map. There exist
some special cases of completely positive maps. For example,
\begin{equation}\label{HH6}
\rho\to -\varepsilon\rho+\frac{1+\varepsilon}{N}\,\rho
\end{equation}
differs from the depolarizing map by replacing the unity operator by
the density operator. Another map reads
\begin{equation}\label{HH7}
\rho\to\frac{1-\mbox{diag}\,\rho}{N}\,.
\end{equation}
The decoherence map (phase damping map) of the kind
\begin{equation}\label{HH8}
\rho_{ij}\to\left\{\begin{array}{clcr}\rho_{ij},&i=j\\
\lambda\rho_{ij},&i\neq j,\end{array}\right.\end{equation}
where $|\lambda|<1$ provides contractive map with uniform change of
off-diagonal matrix elements of the density matrix.

Let us discuss the property of tomogram of bipartite system with
density matrix $\rho_{12}$. If the density matrix is separable, than
the depolarizing map of the second subsystem provides the following
density matrix
\begin{equation}\label{HH9}
\rho_{12}\to\rho_\varepsilon =-\varepsilon\rho_{12}+\frac{1+
\varepsilon}{N_2}\,\underline{\rho^{(1)}}\otimes 1_2,
\end{equation}
where
\begin{equation}\label{HH10}
\underline{\rho^{(1)}}=\mbox{Tr}_2(\rho_{12})
\end{equation}
and $1_2$ is the $N_2$-dimensional unity matrix. Then one has the
property of unitary spin tomogram
\begin{equation}\label{HH11}
w_\varepsilon(m_1,m_2,g^{(n)})=-\varepsilon w_{12}(m_1,m_2,g^{(n)})
+\frac{1+\varepsilon}{N_2}\,\underline w(m_1,m_2,g^{(n)}),
\end{equation}
where $g^{(n)}$ is matrix of $U\Big((2j_1+1)(2j_2+1)\Big)$ unitary
transform;\\ $w_\varepsilon (m_1,m_2,g^{(n)})$  is the tomogram of
transformed density matrix of bipartite system;\\
$\underline{w}(m_1,m_2,g^{(n)})$ is the unitary spin tomogram of
tensor product of partial trace  $\underline{\rho^{(1)}}$ over the
second subsystem's coordinates of the density matrix $\rho_{12}$ and
unity operator $1_2$;\\ $w_{12}(m_1,m_2,g^{(n)})$ is the unitary spin
tomogram of the state with density matrix $\rho_{12}$.

The criterion of separability means
\begin{equation}\label{HH12}
\sum_{m_1=-j_1}^{j_1}\sum_{m_2=-j_2}^{j_2}\left|\frac{1+\varepsilon}{2j_2+1}
\underline{w}(m_1,m_2,g^{(n)})-\varepsilon w_{12}(m_1,m_2,g^{(n)})
\right|=1
\end{equation}
for arbitrary $g^{(n)}$ and $\varepsilon$.

For Werner states $\rho_W$, the tomogram of transformed state (in this
case, it means that $p\to -\varepsilon p$) is related to the
initial-state tomogram  $w_W$
\begin{equation}\label{HH13}
w_\varepsilon(m_1,m_2,g^{(n)})=-\varepsilon w_{12}(m_1,m_2,g^{(n)})
+\frac{1+\varepsilon}{4}\,.
\end{equation}
The criterion of separability yields
\begin{equation}\label{HH14}
\sum_{m_1,m_2=-1/2}^{1/2}\left|\frac{1+\varepsilon}{4}
-\varepsilon w_W(m_1,m_2,g^{(n)})\right|=1.
\end{equation}
Equality~(\ref{HH14}) takes place for arbitrary $g^{(n)}$ and
$\varepsilon$ only for $|p|\leq 1/3$. For $p>1/3$, the above sum
depends on $g^{(n)}$ and $\varepsilon$ and it is larger than one.

It is obvious if one calculates the tomogram using the element of the
unitary group of the form
\begin{equation}\label{H15}
g_0^{(n)} =\left(
\begin{array}{cccc}
0& 0 & 0 & 1 \\ 0 & 1 & 0 & 0 \\ 0 & 0&1 & 0 \\ 1 & 0 & 0 &
0\end{array}
\right).
\end{equation}
At this point, the sum (\ref{HH14}) reads
\begin{equation}\label{HH16}
\sum_{m_1,m_2=-1/2}^{1/2}\left|\frac{1+\varepsilon}{4}
-\varepsilon w_W(m_1,m_2,g^{(n)})\right|=3\left|\frac{1+\varepsilon p}{4}
\right|+\left|\frac{1-3p\varepsilon}{4}\right|.
\end{equation}
One can see that this sum equals to one independently on the value of
parameter $|\varepsilon|\leq 1$ only for values $|p|\leq 1/3\,.$ For
$p=1$, the maximum value of the sum equals
$2=(1+3\varepsilon)/2~(\varepsilon=1)$. This value can characterize
the degree of entanglement of Werner state.

We have introduced positive nonlinear map of density matrix which is
purification map. The purification map can be combined with
contractive maps discussed. The tomographic-probability distributions
under discussion can be completely described by their characteristic
functions. This means that the relation of tomogram property to
entanglement can be formulated in terms of the properties of
characteristic functions.

One can also check the criterion using example of two-qutrite pure
entangled state with wave function
\begin{equation}\label{HH17}
\mid \psi\rangle=\frac {1}{\sqrt 3}\,\sum_{m=-1}^1\mid
u_m\rangle\mid v_m\rangle.
\end{equation}
The sum defining the criterion of separability for specific $U(9)$
transform $g_0^{(n)}$ which is diagonalizing the hermitian matrix
$L_\varepsilon\mid\psi\rangle\langle\psi\mid$ reads
\begin{equation}\label{HH18}
F(\varepsilon,g_0^{(n)})=8\left|\frac{1+\varepsilon}{9}\right|+
\left|\frac{1-8\varepsilon}{9}\right|.
\end{equation}
For $1/2>\varepsilon>1/8$, this sum is larger than one, that means
that the state is entangled. For $\varepsilon=1/2$, the function has
maximum and it is equal to 5/3.

The entanglement of the considered state can be detected using partial
transposition criterion too.

For the case of pure entangled state of two-qutrite system with the
wave function
\begin{equation}\label{HH19}
\mid\Phi\rangle=\frac{1}{\sqrt 2}\,\Big(\mid u_1\rangle\mid v_1\rangle+
\mid u_0\rangle\mid v_0\rangle\Big),
\end{equation}
in which the states with spin projections $m=-1$ do not participate,
the partial transpose criterion does also detect entanglement. Our
criterion yields for specific $U(9)$ transform $g_0^{(n)}$, which
diagonalizes the hermitian matrix $L_\varepsilon\mid
\Phi\rangle\langle\Phi\mid$
the following expression for the function $F(\varepsilon,g_0^{(n)})$,
which reads
\begin{equation}\label{HH20}
F(\varepsilon,g_0^{(n)})=5\,\frac{|1+\varepsilon|}{6}+\frac{|1
-5\varepsilon|}{6}\,.\end{equation}

The function takes maximum value for $\varepsilon=1/2$ that equals to
3/2. This value is smaller than 5/3 of the previous case. It
corresponds to our intuition that the superposition of three   product
states of two qutrite system is more entangled than the superposition
of only two such product states.

The criterion can be extended to multipartite spin system.

We have to apply for $n$-partite system the transform of the density
matrix $\rho$ of the form
\begin{equation}\label{HH21}
L_{\vec\varepsilon}=L_{\varepsilon_1}^{(1)}\otimes
L_{\varepsilon_2}^{(2)}\otimes \ldots\otimes
L_{\varepsilon_n}^{(n)},\end{equation} where the transform
$L_{\varepsilon_k}^{(k)}$ acts as depolarizing map on the $k$th
subsystem. If the state is separable
\begin{equation}\label{HH22}
\rho=\sum_k p_k\rho_k^{(1)}\otimes
\rho_k^{(2)}\otimes \ldots\otimes \rho_k^{(n)},
\quad \sum_kp_k=1,\quad p_k\geq 0,
\end{equation}
each of the terms $\rho_k^{(j)}~(j=1,2,\ldots ,n)$ in the tensor
product is replaced by the term
\begin{equation}\label{HH23}
\rho_k^{(j)}\to -\varepsilon_j\rho_k^{(j)}+\frac{1
+\varepsilon_j}{N_j}\,1_j.
\end{equation}
This means that the transformed density matrix reads
\begin{equation}\label{HH24}
\rho\to L_{\vec\varepsilon}\rho=\sum_kp_k
\left[\prod_{j=1}^n\otimes\left(-\varepsilon\rho_k^{(j)}+\frac{1
+\varepsilon_j}{N_j}\,1_j\right)\right].
\end{equation}
The unitary spin tomogram of the transformed density matrix takes the
form  $(\vec\varepsilon=\varepsilon_1,\varepsilon_2,\ldots
,\varepsilon_n)$
\begin{equation}\label{HH25}
w_{\vec\varepsilon}(m_1,m_2,\ldots ,m_n,g^{(N)})=\sum_kp_k
w_{pr}^{(k)}(m_1,m_2,\ldots ,m_n,g^{(N)},\vec\varepsilon),
\end{equation}
where $N=\prod_{s=1}^n(2j_s+1)$ and element $g^{(N)}$ is the unitary
matrix in $N$-dimensional space. The tomogram
$w_{pr}^{(k)}(m_1,m_2,\ldots ,m_n,g^{(N)},\vec\varepsilon)$ is the
joint probability distribution of spin projections
$m_s=-j_s,-j_s+1,\ldots ,j_s$, which depends on the unitary transform
 $g^{(N)}$ in the state with density matrix
\begin{equation}\label{HH26}
\rho_k=\prod_{s=1}^n\otimes\left(-\varepsilon_s\rho_k^{(s)}+\frac{1
+\varepsilon_s}{N_s}\,1_s\right).
\end{equation}
For the elements $$g_{pr}^{(N)}=\prod_{s=1}^n\otimes u_s(2j_s+1),$$
where $u_s(2j_s+1)$ is unitary matrix, the tomogram (\ref{HH25}) takes
the form of sum of the products
\begin{equation}\label{HH27}
w_{\vec\varepsilon}(m_1,m_2,\ldots
,m_n,g_{pr}^{(N)})=\sum_kp_k\prod_{s=1}^n w_k\Big(m_s,u_s
(2j_s+1),\varepsilon_s\Big),
\end{equation}
with $w_k(m_s,u_s(2j_s+1),\varepsilon_s)$ being the unitary spin
tomograms of the $s$th spin subsystem with transformed density matrix
$L_{\varepsilon_s}\rho_k^{(s)}$. If one uses as the matrix
$u_s(2j_s+1)$, the matrix of unitary irreducible representation of the
$SU(2)$ group, the tomogram $w_k$ depends only on the two parameters
defining the point on the sphere $S^2$.

For a separable state of the multipartite system, one has
\begin{equation}\label{HH28}
\sum_{m_1,\ldots,m_n}\left|w_{\vec\varepsilon}(m_1,m_2,
\ldots ,m_n,g^{(N)})\right|=1
\end{equation}
for all elements $g^{(N)}$ and all parameters $\vec\varepsilon$.

For entangled state, there can be some values of parameters
$\vec\varepsilon$ and group elements $g^{(N)}$ for which the sum is
larger than one.

\section{Conclusions}

We summarize the results of the paper.

The notion of entangled states (first discussed by
Schr\"odinger~[4,~51]) has attracted a lot of efforts to find a
criterion and quantitative characteristics of entanglement. A
criterion based on partial transpose transform of subsystem density
matrix (complex conjugation of the subsystem density matrix or its
time reverse) provides the necessary and sufficient condition of
separability of the system of two qubits and qubit-qutritt
system~[52].
The phase-space representation of the quantum states and time reverse
transform (change of the signs of the subsystem momenta) of the Wigner
function in the case of Gaussian state was applied to study the
separability and entanglement of photon states in [13].
Recently it was pointed out that the tomographic approach of
reconstructing the Wigner function of quantum state~[43--45]
can be developed to consider the positive probability distribution
(tomogram) as an alternative to density matrix (or wave function)
because the complete set of tomograms contains the complete
information on the quantum state~[42].
This representation (called probability representation) was
constructed also for spin states including a bipartite system of two
spins. Up to now the problem of entanglement was not discussed in the
tomographic representation. Some remarks on tomograms and entanglement
of photon states in the process of Raman scattering were done in [53].
The tomographic approach has the advantage of dealing with positive
probabilities and one deals with standard probability distributions
which are positive and normalized.

We studied the properties of separable and entangled state of
multipartite system using the tomographic probability distributions.
The positive and completely positive maps of density matrices~[39,~54]
induce specific properties of the tomograms. The properties of the
positive maps were studied in [55].
We formulated necessary and sufficient conditions of separability and
entanglement of multipartite systems in terms of properties of the
quantum tomogram. Since the tomograms were shown~[36]
to be related to the star-product quantization
procedure~[56],
we discuss entanglement and separability properties in terms of
generic operator symbols. The tomographic symbols of generic spin
operators were studied in [36].
Then we focused on properties of entanglement and separability of a
bipartite system using spin tomograms ($SU(2)$-tomograms) and
tomograms of the $U(N)$-group.

The idea of the approach suggested can be summarized as following.

The positive but not completely positive linear maps of a subsystem
density matrix do preserve the positivity of separable density matrix
of the composite system. These maps contain also maps which do not
preserve the positivity of the initial density matrix of an entangled
state for the composite system. It means that the set of all linear
positive maps of the subsystem density matrix (this set is semigroup)
creates from the initial entangled positive density matrix of
composite system a set of hermitian matrices including the matrices
with negative eigenvalues. To detect the entanglement we use the
tomographic symbols of the obtained hermitian matrices.The tomographic
symbols of state density matrices (state tomograms)are standard
probabilities.In view of this the tomographic symbols of the obtained
hermitian matrices corresponding to initial separable state preserve
all the properties of the probability representation including
positivity and normalization. But in case of entangled state the
tomographic symbols of the obtained hermitian matrices can take
negative values.The different behaviour of tomograms of separable and
entangled states of composite systems under action of the semigroup of
positive maps of the subsystem density matrix provides the tomographic
criterion of the separability.

To conclude, we point out the main result of the work.

We found the criterion of separability which is given by
equation~(\ref{U21}). The criterion is valid for multiparticle spin
system. The criterion can be called ``tomographic criterion'' of
separability. The tomographic criterion can be considered also for
symplectic tomograms of multimode photon states. The condition of
separability is sufficient because there always exists a unitary group
element by means of which any hermitian matrix can be diagonalized.It
means that tomographic symbol of nonpositive hermitian matrix has
nonpositive values for some unitary group parameters. The suggested
criterion is connected with properties of the constructed function
(\ref{U21}) which for given density matrix depends on unitary group
parameters $g$ and the parameters of positive map semigroup $L$. For
separable density matrix the dependence on unitary group parameters
and the semigroup parameters disappears and the function becomes
constant equal to unity. For entangled states the function differs
from unity and depends on both group and semigroup parameters. The
suggested criterion can be considered as some complementary test of
separability together with other criteria available in the literature
(see, for example, [38,~52]).
We point out that suggested criterion differs from available usual
ones by the kind of the necessary numerical calculations. To apply
this criterion one needs to calculate the sum of moduli of diagonal
matrix elements of product of three matrices. One of the matrices is
hermitian and two others are unitary ones. This procedure does not
need the calculation of the eigenvalues of a matrix. The structure of
positive (including not completely positive) map semigroup with
elements $L$ needs extra investigation (see, for example, [57]).
We found also a test of entanglement based on the property of unitary
spin tomogram.

The discussed purification map can be applied to find new quantum
evolution equations in addition to known ones~[58--61]. The
application of different forms of positive maps~[55,~62] and
supermatrix representation of the maps~[63,~64] are useful for better
understanding of the computations. Entanglement phenomena can be
considered using symbols of density matrix of different kinds, e.g.,
particular quasidistributions~[65]
as well as tomographic symbols~[36]. The difference of symbols of
entangled and separable density operators for different schemes of the
star-product quantization needs further investigations as well as test
of entanglement of some generalizations of Werner state~[66,~40] in
multipartite case. A relation of tomographic approach to different
positive maps~[67] should be investigated. The tomographic symbols are
analytic in group parameters. This can be used to find extrema of
tomograms which give information on degree of entanglement.

\section*{Acknowledgments}

V.~I.~M. and E.~C.~G.~S. thank Dipartimento di Scienze Fisiche,
Universit\'a ``Federico~II'' di Napoli and Istitito Nazionale di
Fisica Nucleare, Sezione di Napoli for kind hospitality. V.~I.~M. is
grateful to the Russian Foundation for Basic Research for partial
support under Project~No.~01-02-17745.

\end{article}

\begin{thebibliography}{99}
\bibitem{Dirac}
[1]~P.~A.~M.~Dirac 1958 ``The Principles of Quantum Mechanics''
(Oxford: Pergamon)

\bibitem{Landau}
[2]~L. D. Landau 1927 {\it Z. Phys} {\bf 45} 430

\bibitem{vonNeumann}
[3]~J. von Neumann 1932 ``Mathematische Grudlagen der
Quantenmechanik'' (Berlin: Springer); Nov.~1927 {\it G\"{o}ttingenische
Nachrichten} {\bf 11} S245

\bibitem{Schroedinger}
[4]~E. Schr\"odinger 1935 {\it Naturwissenschaften} {\bf 23} 807; 823;
844

\bibitem{Schr26}
[5]~E. Schr\"odinger 1926 {\it Ann. d. Phys.} Lpz {\bf 79} 489

\bibitem{MMSZ-J.Phys.A}
[6]~V. I. Man'ko, G. Marmo, E. C. G. Sudarshan and F. Zaccaria 1999
{\it J. Russ. Laser Res.} {\bf 20} 421; 2002 {\it J. Phys. A: Math.
Gen.} {\bf 35} 7173

\bibitem{Plenio}
[7]~M. Horodecki, P. Horodecki and R. Horodecki 1996 {\it Phys. Lett.}
A {\bf 223} 1

\bibitem{Vedral}
[8]~S. Hill and W. K. Wootters 1997 {\it Phys. Rev. Lett.}  {\bf 78}
5022  \newline W. K. Wootters 1998 {\it Phys. Rev. Lett.} {\bf 80}
2245

\bibitem{Teich}
[9]~K. Zyczkowski, P. Horodecki, A. Sanpera and M. Lewenstein 1998
{\it Phys. Rev.} A {\bf 58} 883

\bibitem{Uhlman}
[10]~S. Popescu and D. Rohrlich 1997 {\it Phys. Rev.} A {\bf 56} R3319

\bibitem{Dodonov}
[11]~S. Abe and A. K. Rajagopal 2002 {\it Physica} A {\bf 289} 157

\bibitem{entropy}
[12]~C. H. Bennett, D. P. Di Vincenzo, J. A. Smolin and W. L. Wootters
1996 {\it Phys. Rev.} A {\bf 54} 3824

\bibitem{concurrence}
[13]~R. Simon 2002 {\it Phys. Rev. Lett.} {\bf 84} 2726

\bibitem{Bar89}
[14]~S. M. Barnett and S. J. D. Phoenix  1989 {\it Phys. Rev.} A {\bf
40} 2404; 1991 {ibid} {\bf 44} 535

\bibitem{Mann95}
[15]~A. Mann, B. C. Sanders and W. J. Munro 1995 {\it Phys. Rev.} A
{\bf 51} 989

\bibitem{Ben96}
[16]~C. H. Bennet, H. J. Bernstein, S. Popescu and B. Schumacher 1996
{\it Phys. Rev.} A {\bf 53} 2046

\bibitem{Ved98}
[17]~V. Vedral and M. B. Plenio 1998 {\it Phys. Rev.} A {\bf 57} 1619

\bibitem{Engl198}
[18]~B.-G. Englert, M. L\"{o}ffler, O. Benson, B. Varcoe, M. Weidinger and
H. Walter  1998 {\it Fortschr. Phys.} {\bf 46} 897

\bibitem{3Hor00}
[19]~M. Horodecki, P. Horodecki and R. Horodecki 2000 {\it Phys. Rev.
Lett.} {\bf 84} 2014

\bibitem{Par00}
[20]~S. Parker, S. Bose and M. B. Plenio 2000 {\it Phys. Rev.} A {\bf
61} 032305

\bibitem{Benn01}
[21]~C. H. Bennet, S. Popescu, D. Rohrlich, J. A. Smolin and A. V.
Thapliyal 2001 {\it Phys. Rev.} A {\bf 63} 012307

\bibitem{Pia01}
[22]~K. Pi\c{a}tek and W. Leo\'nski 2001 {\it J. Phys. A: Math. Gen.}
{\bf 34} 4951

\bibitem{Aud01}
[23]~K. Audenaert, J. Esert, E. Jan\'{e}, M. B. Plenio, S. Virmani and R.
B. De Moor 2001 {\it Phys. Rev. Lett.} {\bf 87} 217902

\bibitem{Fur98}
[24]~K. Furuya, M. C. Nemes and G. Q. Pellegrino 1998 {\it Phys. Rev.
Lett.} {\bf 80} 5524

\bibitem{Zyc98}
[25]~K. \v{Z}yczkowski, P. Horofecki, A. Savpera and M. Lewenstein
1998 {\it Phys. Rev.} A {\bf 58} 883

\bibitem{Mun01}
[26]~W. J. Munro, D. F. V. James, A. G. White and P. G. Kwait 2001
{\it Phys. Rev. } A {\bf 64} 030302

\bibitem{VidWer02}
[27]~G. Vidal and R. F. Werner 2002 {\it Phys. Rev.} A {\bf 65} 032314

\bibitem{Goff00}
[28]~F. Coffman, J. Kundu and W. K. Wootters 2000 {\it Phys. Rev.} A
{\bf 61} 052306

\bibitem{Badz02}
[29]~P. Badziag, P. Deuar, M. Horodecki, P. Horodecki and R. Horodecki
2002 {\it J. Mod. Opt.} {\bf 49} 1289

\bibitem{AlDod}
[30]~M. A. Andreata, A. V. Dodonov and V. V. Dodonov 2002 {\it J.
Russ. Laser Res.} {\bf 23} 531

\bibitem{Dod}
[31]~V. V. Dodonov and V. I. Man'ko 1997 {\it Phys. Lett.} A {\bf 229}
335

\bibitem{Olga}
[32]~Olga Man'ko and V. I. Man'ko 1997 {\it JETP} {\bf 85} 430

\bibitem{Klim}
[33]~A. B. Klimov, O. V. Man'ko, V. I. Man'ko, Yu. F. Smirnov and V.
N. Tolstoy 2002 {\it J. Phys. A: Math. Gen.} {\bf 35} 6101

\bibitem{Andreev}
[34]~V. A. Andreev and V. I. Man'ko 1998 {\it JETP} {\bf 87} 239

\bibitem{Andreev-Olga}
[35]~V. I. Man'ko and S. S. Safonov 1998 {\it Yad. Fiz.} {\bf 61} 658

\bibitem{Marmo}
[36]~O. V. Man'ko, V. I. Man'ko and G. Marmo 2000 {\it Phys Scr.} {\bf
62} 446; 2002 {\it J. Phys. A: Math. Gen.} {\bf 35} 699

\bibitem{DeCastro}
[37]~V. V. Dodonov, A. S. M. De Castro and S. S. Misrahi 2002 {\it
Phys. Lett.} A {\bf 296} 73\newline A. S. M. De Castro and V. V.
Dodonov 2003 {\it J. Russ. Laser Res.} {\bf 23} 93; 2003 {\it J. Opt.
B: Quantum Semiclass. Opt.} {\bf 5} S593

\bibitem{J.Math.Phys:SI-Sept.2002}
[38]~Special Issue on Entanglement 2002 {\it J. Math. Phys.} {\bf 43}
No.~9

\bibitem{Sud61}
[39]~E. C. G. Sudarshan, P. M. Mathews and J. Rau 1961 {\it Phys.
Rev.} {\bf 121} 920

\bibitem{geom}
[40]~E. C. G. Sudarshan and A. Shaji 2003 ``Structure and
parametrization of stochastic maps of density matrix'' ArXiv
quant-ph/0205051 v2; 2003 {\it J. Phys. A: Math. Gen.} {\bf 36} (in
press)

\bibitem{Peres}
[41]~A. Peres 1996 {\it Phys. Rev. Lett.} {\bf 77} 1413

\bibitem{JOB}
[42]~S. Mancini, V. I. Man'ko and P. Tombesi 1996 {\it Phys. Lett.} A
{\bf 213} 1; 1997 {\it Found. Phys.} {\bf 27} 801

\bibitem{14}
[43]~J. Bertrand and P. Bertrand 1987 {\it Found. Phys.} {\bf 17} 397

\bibitem{15}
[44]~K. Vogel and H. Risken 1989 {\it Phys. Rev.} A {\bf 40} 2847

\bibitem{16}
[45]~S. Mancini, V. I. Man'ko and P. Tombesi 1995 {\it Quantum
Semiclass. Opt.} {\bf 7} 615 \newline G. M. D'Ariano, S. Mancini, V.
I. Man'ko and P. Tombesi 1996 {\it Quantum Semiclass. Opt.} {\bf 8}
1017

\bibitem{MendesJPA}
[46]~M. A. Man'ko, V. I. Man'ko and R. V. Mendes 2001 {\it J. Phys. A:
Math. Gen.} {\bf 24} 8321

\bibitem{JOBnew}
[47]~V. I. Man'ko, G. Marmo, E. C. G. Sudarshan and F. Zaccaria 2003
``Entanglement in probability representation of quantum states and
tomographic criterion of separability'' {\it J. Opt. B: Quantum
Semiclass. Opt.} (in press); 2003 {\it J. Russ. Laser Res.} {\bf 24}
507

\bibitem{Tino-book}
[48]~V. I. Man'ko, G. Marmo, E. C. G. Sudarshan and F. Zaccaria 2000
``Inner composition law of pure-spin states'' in ``Spin-Statistics
Connection and Commutation Relations'' R. C. Hilborn and G. M.
Tino~(eds.) {\it AIP Conference Proceedings} {\bf 545} 92

\bibitem{Holevo}
[49]~A. S. Holevo 1999 {\it Russ. Math. Surveys} {\bf 53} 1295

\bibitem{Shor}
[50]~P. W. Shor 2002 {\it J. Math. Phys.} {\bf 43} 4334

\bibitem{Schrod-Cambridge}
[51]~E. Schr\"odinger 1935 {\it Proc. Cambridge Philos. Soc.} {\bf 31}
555

\bibitem{Slater}
[52]~P. B. Slater 2003 {\it J. Opt. B: Quantum Semiclass. Opt.} {\bf
5} S691

\bibitem{Olga-JOBSI}
[53]~S. V. Kuznetsov, O. V. Man'ko and N. V. Tcherniega 2003 {\it J.
Opt. B: Quantum Semiclass. Opt.} {\bf 6} S503

\bibitem{Stinespr}
[54]~W. F. Stinespring 1955 {\it Proc. Amer. Math. Soc.} {\bf 6} 211

\bibitem{woron}
[55]~L. C. Woronowicz 1976 {\it Rep. Math. Phys.} {\bf 10} 165

\bibitem{Fronsdal}
[56]~F. Bayen, M. Flato, C. Fronsdal, A. Lichnerovicz and D.
Sternheimer 1975 {\it Lett. Math. Phys.} {\bf 1} 521

\bibitem{King}
[57]~C. King 2002 ``The capacity of the quantum depolarizing channel''
ArXiv quant-ph/0204172 v2

\bibitem{Moyal}
[58]~J. E. Moyal 1949 {\it Proc. Cambridge Philos. Soc.} {\bf 45} 99

\bibitem{Kosak}
[59]~A. Kossakovski 1972 {\it Rep. Math. Phys.} {\bf 3} 247

\bibitem{Lind}
[60]~G. Lindblad 1976 {\it Comm. Math. Phys.} {\bf 48} 119

\bibitem{Gorini}
[61]~V. Gorini, A. Kossakovski and E. C. G. Sudarshan 1978 {\it Rep.
Math. Phys.} {\bf 18} 149

\bibitem{Kraus}
[62]~K. Kraus 1973 {\it Ann. Phys.} NY {\bf 64} 311

\bibitem{Choi}
[63]~M. D. Choi 1975 {\it Lin. Alg. Appl.} {\bf 10} 285; 1970 {\it
Canadian J. Math.} {\bf 24} 520; 1976 {\it Illinois J. Math.} {\bf 48}
119

\bibitem{Havel}
[64]~T. F. Havel 2003 {\it J. Math. Phys.} {\bf 44} 534

\bibitem{Sud63}
[65]~E. C. G. Sudarshan 1963 {\it Phys. Rev. Lett.} {\bf 10} 277\\ C.
L. Mehta and E. C. G. Sudarshan 1965 {\it Phys. Rev.} B {\bf 138} 274

\bibitem{Werner}
[66]~R. Werner 1989 {\it Phys. Rev.} A {\bf 40} 4277

\bibitem{AAA}
[67]~A. Kossakowski 2003 ``A class of linear positive maps in matrix
algebras'' ArXiv quant-ph/0307132 v1

\end{thebibliography}
\end{document}